\pdfoutput=1

\documentclass{raa04}
\usepackage{graphicx,times}
\usepackage{natbib}
\usepackage{amssymb,amsmath}
\usepackage[unicode]{hyperref}

\bibpunct{(}{)}{;}{a}{}{,}

\begin{document}

\title{Determining the type of orbits in the central regions of barred galaxies}

\volnopage{{\bf 2016} Vol.\ {\bf 16} No. {\bf 2}, ~ 26 (18pp)~
 {\small doi: 10.1088/1674--4527/16/2/026}}
      \setcounter{page}{1}

\author{Euaggelos E. Zotos
\and Nicolaos D. Caranicolas}

\institute{Department of Physics,
Section of Astrophysics, Astronomy and Mechanics,
Aristotle University of Thessaloniki
GR-541 24, Thessaloniki, Greece;
{\it evzotos@physics.auth.gr}\\
\vs \no
   {\small Received 2015 April 17; accepted 2015 July 15}
}

\abstract{ We use a simple dynamical model which consists of a
harmonic oscillator and a spherical component, in order to
investigate the regular or chaotic character of orbits in a barred
galaxy with a central spherically symmetric nucleus. Our aim is to
explore how the basic parameters of the galactic system influence
the nature of orbits, by computing in each case the percentage of
chaotic orbits, as well as the percentages of different types of
regular orbits. We also give emphasis to the types of regular orbits
that support either the formation of nuclear rings or the barred
structure of the galaxy. We provide evidence that the traditional
$x_1$ orbital family does not always dominate in barred galaxy
models since we found several other types of resonant orbits which
can also support the barred structure. We also found that sparse
enough nuclei, fast rotating bars and high energy models can support
the galactic bars. On the other hand, weak bars, dense central
nuclei, slow{ly} rotating bars and low energy models favor the
formation of nuclear rings. We also compare our results with
previous related work.
 }
\authorrunning{E. E. Zotos \& N. D. Caranicolas}
\titlerunning{Determining the Type of Orbits in the Central Regions of Barred Galaxies}
\maketitle

\section{Introduction}
\label{intro}

It is well known to astronomers that axial symmetry in galaxies is
only a first {approximation}. In essence, galaxies exhibit
deviation from axial symmetry, which can be very small or more
extended. In the latter category, we may include the case of
barred galaxies. Observations indicate that a large percentage of
disk galaxies, about 65\%, show bar-like formations
\citep[e.g.,][]{Ee00,SRSS03}. Observations also show that barred
galaxies may display different characteristics. There are galaxies
with a prominent barred structure and also galaxies with faint
weak bars. Moreover, there are also barred galaxies with massive
and less massive bulges. In some cases, the formation of the
central bulges is the result caused by dynamical instabilities{ in
the disk} \citep{KK04}.

In fact, bars are non-axisymmetric structures that can redistribute
the angular momentum of a galaxy, thus favoring the transport
of gas through them to the inner regions where it may trigger star
formation and play an important role in the evolution of these
stellar systems \citep[e.g.,][]{MCA04,SVRTT05}. In this way,
considerable changes occur in the morphology and structure of barred
galaxies. Such a change is{ exhibited by} {a} significant increase in the mass of
the bulge in galaxies with bars \citep[e.g.,][]{FM93,MJ07}.

An important and striking phenomenon in barred galaxies is
associated with nuclear rings, which are active sites of new star
formation \citep[e.g.,][]{KBHSd95,MKVR08,Se10,HMLHOW11}. Some scientists have the impression that the formation of
nuclear rings is due to the effect of {the }non axially
symmetric potential of the bar in {an} {area with }plenty of
interstellar gas. {A k}ey role in this
mechanism is played by the torque of the bar, which causes the gas
to form the nuclear rings \citep{KSSYT12}. Observations show that
the rate of star formation in the nuclear rings {is }not only
different in several types of barred galaxies but also varies
significantly with time \citep[e.g.,][]{BTBC00,BHJKS02,CKB10}.

The formation and evolution of dust lanes and nuclear rings have
been extensively studied using numerical simulations
\citep[e.g.,][]{PST95,EG97,MTSS02,RT03,TAJ09}. The formation of
nuclear rings from the resonant interaction of gas with the
potential of the bar appears not to be consistent with recent
studies indicating that these formations are probably due to a
different mechanism \citep[e.g.,][]{KSK12}. According to this
mechanism, there is a centrifugal barrier which cannot be overcome
by the inflowing gas. This barrier is responsible for the formation
of the nuclear rings. Finally, recent research reveals that
the more massive the bar{ is,} the smaller the formed
nuclear rings{ are}. Here we should mention that the observational
data justify the above results \citep{CKB10}.

Over {recent} decades, a huge amount of research
work has been devoted {to} understanding the orbital
structure in barred galaxy models
\citep[e.g.,][]{ABMP83,P84,CDFP90,A92,P96,KC96,OP98,PMM04}. The
reader can find more information about the dynamics of barred
galaxies in the reviews by \citet{A84,CG89,SW93}. We would like to
point out that all the above-mentioned references on the
dynamics of barred galaxies are exemplary rather than exhaustive.
However, we should like to discuss briefly some of the recent papers
on this subject. \citet{SPA02a} conducted an extensive investigation
regarding the stability and morphology of both {two-dimensional (}2D{)} and {three-dimensional (}3D{)} periodic
orbits in a fiducial model representative of a barred galaxy. The
work was continued in the same vein in \citet{SPA02b}, where the
influence of the system's parameters on the 3D periodic orbits was
revealed. Moreover, \citet{KP05} presented evidence that in
{2D} models with sufficiently large bar axial ratios,
stable orbits having propeller shapes play a dominant role {in} the
bar structure. \citet{MA11} estimated the fraction of chaotic and
regular orbits in both {2D} and {3D} potentials by
computing several sets of initial conditions and studying how these
fractions evolve when the energy and also basic parameter{s} of the
model, such as the mass, size and pattern speed of the bar{,} vary.
Computing the statistical distributions of sums of position
coordinates{,} \citet{BMA12} quantified weak and strong chaotic orbits
in 2D and 3D barred galaxy models. A time-dependent barred galaxy
model was utilized in \citet{MBS13} in order to explore the
interplay between chaotic and regular behavior of star orbits when
the parameters of the model evolve. Finally, in \cite{CP07,CZ10}{,} we
conducted an investigation {only }regarding the issue of regular vs
chaotic orbits in simple barred spiral potentials. In the present
paper on the other hand, we shall try to contribute to this active
field by classifying ordered orbits into different regular families
and monitor how their rates evolve when basic quantities of the
system vary.

\citet{LS92}, in a thorough pioneer{ing} study, analyzed the
orbital content in the coordinate planes of triaxial potentials
{and} also in the meridional plane of axially symmetric model
potentials, focusing on regular families. Few years later,
\citet{CA98} developed a method based on the analysis of the
Fourier spectrum of the orbits which can {not only }distinguish
between regular and chaotic orbits, but also between loop, box and
other resonant orbits either in {2D} or {3D} potentials. This
spectral method was improved and applied in \citet{MCW05} in order
to identify the different kinds of regular orbits in a
self-consistent triaxial model. The same code was improved even
further in \citet{ZC13}, when the influence of the central nucleus
and of the isolated integrals of motion (angular momentum and
energy) on the percentages of orbits in the meridional plane of an
axisymmetric galactic model composed of a disk and a spherical
nucleus were investigated. In two recent papers{,} \citet{CZ13}
and \citet{ZCar13}, analytical dynamical models describing the
motion of stars {in }both disk and elliptical galaxies containing
dark matter were used in order to investigate how the presence and
the amount of dark matter influences the regular or chaotic nature
of orbits as well as the behavior of the different families of
resonant orbits.

Taking into account all the above, there is no doubt that
knowing the overall orbital structure in the central regions of
barred galaxies is an issue of paramount importance. For this
reason, we decided to use a simple model that describes local motion
near the central area of a barred galaxy. Our aim is to investigate
the regular or chaotic character of motion and to study how the
different families of orbits are affected by varying the physical
quantities entering the model. Here we must point out that
the present article belongs to a series of papers
\citep{ZC13,CZ13,ZCar13,ZCar14} that have as {their }main
objective the orbit classification (not only regular versus chaotic
but also separating regular orbits into different regular families)
in different galactic gravitational potentials. Thus, we decided to
follow a similar structure and {apply }the same numerical approach to all of
them.

The structure of the present paper is as follows: In
Section~\ref{galmod} we present a detailed description of the
properties of our gravitational galactic model. All the different
computational methods used in order to determine the character of
orbits are described in Section~\ref{compmeth}. In the following{,}
Section{ 4}, we explore how the basic parameters {involved in} the
dynamical system influence the percentages of all types of orbits
and which of them support either{ a} bar or ring structure. Our article
ends with Section~\ref{disc}, where the conclusions and the
discussions of this research are presented.

\section{Properties of the galactic model}
\label{galmod}

The total gravitational potential $\Phi(x,y)$ consists of two components: the bar potential $\Phi_{\rm b}$ and the central, spherical component $\Phi_{\rm n}$. For the description of properties of the bar we use the following simple harmonic oscillator potential
\begin{equation}
\Phi_{\rm b}(x,y) = \frac{\omega^2}{2}\left(x^2 + \alpha y^2 \right),
\label{Vb}
\end{equation}
where $\alpha$ is a parameter corresponding to the strength of the
bar, while the parameter $\omega$ is used for consistency of the
galactic units.

The spherically symmetric nucleus is modeled by a Plummer potential \citep[e.g.,][]{BT08}
\begin{equation}
\Phi_{\rm n}(x,y) = \frac{-G M_{\rm n}}{\sqrt{x^2 + y^2 + c_{\rm n}^2}}.
\label{Vn}
\end{equation}
Here $G$ is the gravitational constant, while $M_{\rm n}$ and $c_{\rm n}$ are the mass and the scale length of the nucleus, respectively. This potential has been used successfully in the past in order to model and therefore interpret the effects of the central mass component in a galaxy \citep[see, e.g.][]{HN90,HPN93,Z12a,ZC13}. At this point, we must make clear that Equation~(\ref{Vn}) is {neither} intended to represent the potential of a black hole nor that of any other compact object, but just the potential of a dense and massive nucleus{.} {T}herefore, any relativistic effects are out of the scope of this work.

The 2D $\Phi(x,y)$ was chosen because we believe that it is an
approximation for describing local motion near the central parts
of a barred galaxy. Our choice was motivated by two reasons: (i)
the small number of input parameters of Equation~(\ref{Vb}) is an
advantage concerning the performance and speed of the numerical
calculations in comparison with other much more complicated
potentials describing bars (i.e. the Ferrers bar, Ferrers 1877)
 and (ii) it corresponds to{ a} constant density
profile which{,} however, can be assumed when studying local
motion very close to the galactic center, like in our case.
Furthermore, the same potential has been used successfully in many
previous works for modeling the properties of local motion in the
central parts of a galaxy \citep[e.g.,][]{C98,CK98,CP05,Z11,Z12b}.
Note that the density corresponding to the total potential
$\Phi(x,y)${,} on the other hand, is not constant but it
{declines} with increasing distance from the origin due to the
contribution of the spherical nucleus.

The bar rotates counterclockwise at a constant angular velocity
$\Omega_{\rm b}$. Therefore the effective potential is
\begin{equation}
\Phi_{\rm eff}(x,y) = \Phi(x,y) - \frac{1}{2}\Omega_{\rm b}^2 \left(x^2 + y^2 \right).
\label{Veff}
\end{equation}

In our study, we use the well-known system of galactic units, where
the unit of length is 1\,kpc, the unit of mass is $2.325 \times 10^7
{\rm M}_\odot$ and the unit of time is $0.9778 \times 10^8$ yr. The
velocity unit is 10~km~s$^{-1}$, the unit of angular momentum
(per unit mass) is 10~km~kpc$^{-1}$ s$^{-1}$, while $G$ is equal to
unity. The energy unit (per unit mass) is 100 km$^2$s$^{-2}$. In
these units, the values of the involved parameters are: $\omega =
10$, $\alpha = 4$, $M_n = 50$ (corresponding to 1.2 $\times$
$10^{9}$ $M_{\odot}$), $c_{\rm n} = 0.25$ and $\Omega_{\rm b} = 1$.
This set of the values of the parameters defines the Standard Model
(SM).

The equations of motion are described by
\begin{equation}
\ddot{{\vec{r}}} = - {\vec{\nabla}} \Phi_{\rm eff} -
2\left({\vec{\Omega_{\rm b}}} \times \dot{{\vec{r}}} \right),
\label{eqmot0}
\end{equation}
where the term $- 2\left({\vec{\Omega_{\rm b}}} \times
\dot{{\vec{r}}} \right)$ represents the Coriolis force.
Decomposing Equation~(\ref{eqmot0}) into its $x$ and $y$ parts, we
obtain
\begin{equation}
\ddot{x} = - \frac{\partial \Phi_{\rm eff}}{\partial x} + 2\Omega_{\rm b}\dot{y}, \ \ \
\ddot{y} = - \frac{\partial \Phi_{\rm eff}}{\partial y} - 2\Omega_{\rm b}\dot{x},
\label{eqmot}
\end{equation}
where the dot indicates derivative with respect to time.

In the same vein, the equations describing the evolution of a
deviation vector ${\vec{w}} = (\delta x, \delta y, \delta \dot{x},
\delta \dot{y})$ which joins the corresponding phase space points
of two initially nearby orbits, needed for the calculation of
standard chaos indicators (SALI in our case){,} are given by the
following variational equations
\begin{eqnarray}
\dot{(\delta x)} &=& \delta \dot{x}, \ \ \
\dot{(\delta y)} = \delta \dot{y}, \nonumber \\
(\dot{\delta \dot{x}}) &=&
- \frac{\partial^2 \Phi_{\rm eff}}{\partial x^2} \delta x
- \frac{\partial^2 \Phi_{\rm eff}}{\partial x \partial y}\delta y + 2\Omega_{\rm b} \delta \dot{y},
\nonumber \\
(\dot{\delta \dot{y}}) &=&
- \frac{\partial^2 \Phi_{\rm eff}}{\partial y \partial x} \delta x
- \frac{\partial^2 \Phi_{\rm eff}}{\partial y^2}\delta y - 2\Omega_{\rm b} \delta \dot{x}.
\label{vareq}
\end{eqnarray}

Consequently, the corresponding Hamiltonian to the effective potential given in Equation~(\ref{Veff}) reads
\begin{equation}
H_{\rm J} = \frac{1}{2} \left(\dot{x}^2 + \dot{y}^2 \right) + \Phi_{\rm eff}(x,y) = E_{\rm J},
\label{ham}
\end{equation}
where $\dot{x}$ and $\dot{y}$ are momenta per unit mass, conjugate
to $x$ and $y$ respectively, while $E_{\rm J}$ is the numerical
value of the Jacobi integral, which is conserved. Thus, an orbit
with a given value for {its} Jacobi integral is
restricted in its motion to regions in which $E_{\rm J} \leq
\Phi_{\rm eff}$, while all other regions are forbidden{ with respect} to the star.

\section{Computational methods}
\label{compmeth}

For distinguishing between order and chaos in our models we use the
SALI method \citep{S01} (for more details on how the SALI method
works see \citealt{ZC13}). We chose{,} for each value of the free
parameter of the potential, a dense grid of initial conditions in
the $(x,\dot{x})$ phase plane, regularly distributed in the area
allowed by the value of the energy $E_{\rm J}$. In all cases, $y_0 =
0$, while $\dot{y_0}$ is found from the Jacobi integral
(Eq.~(\ref{ham})). The distance between the points of the gird along
the $x$ and $\dot{x}$ directions, or in other words its density, was
calibrated in such a way so {that} every grid contain{s} around {15\,000} orbits. For each
initial condition, we integrated the equations of motion
(\ref{eqmot}) as well as the variational equations (\ref{vareq})
with a double precision Bulirsch-Stoer algorithm
\citep[e.g.,][]{PTVF92}. In all cases, the value of the Jacobi
integral (Eq.~(\ref{ham})) was conserved better than one part in
$10^{-11}$, although for most orbits, it was better than one part in
$10^{-12}$.

All initial conditions of orbits are numerically integrated for
$10^4$ time units which correspond to about $10^{12}$ yr or in{ other}
words to about 100 Hubble times. This vast time of numerical
integration is justified due to the presence of the so called
``sticky orbits"\footnote{A sticky orbit is a chaotic orbit which
behaves as a regular one for a long time period before revealing
its true chaotic nature.}. Therefore, if the integration time is
too short, any chaos indicator will misclassify sticky orbits as
regular ones (see Appendix{ A} for more details and examples). In our
work we decided to integrate all orbits for a time interval of
$10^4$ time units in order to correctly classify sticky orbits
with sticky periods of at least 100 Hubble times. At this point,
it should be clarified that sticky orbits with sticky periods
larger than $10^4$ time units will be counted as regular ones,
since such extremely high sticky periods are completely out of{ the}
scope of our research.

The distinction between order and chaos is only a first step for
interpreting the overall orbital structure of the galactic system.
The second and more important step is the classification of
ordered orbits into different regular families. For this purpose
we use the frequency analysis of \citet{CA98} for categorizing
regular orbits. About thirty years ago{,} \citet{BS82,BS84} proposed
a computational method, dubbed ``spectral dynamics" for the task
of orbit classification, while later on \citet{SN96} and
\citet{CA98} substantially improved this method. \citet{ZC13}
further refined the numerical code for classifying orbits in the
meridional $(R,z)$ plane. The very same algorithm was used in all
the papers of this series{:} \citet{ZC13,CZ13,ZCar13,ZCar14}, as well
as in \citet{MCW05} and \citet{CM12}.

\begin{figure*}

\centering

\includegraphics[width=140mm]{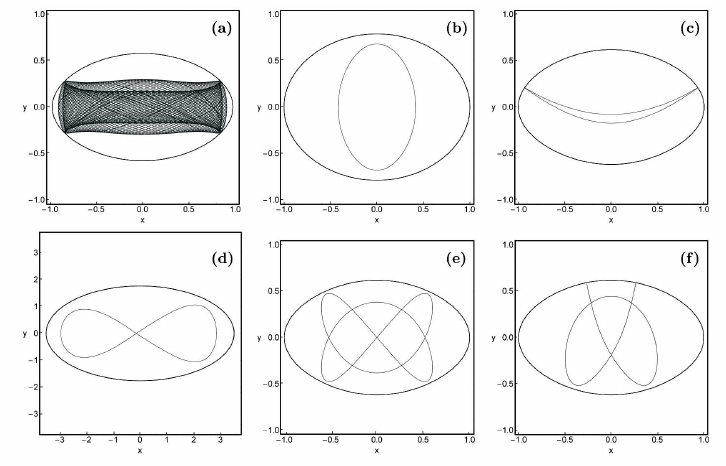}

\caption{\baselineskip 3.6mm Collection of the six basic types of
orbits in our barred galaxy model: (a) box orbit; (b) 1:1 resonant
orbit; (c) 1:2 resonant type a; (d) 1:2 resonant type b
(figure-eight); (e) 2:3 resonant type; (f) 3:4 resonant orbit.}
\label{orbsM}
\end{figure*}

\begin{figure*}

\vs \centering
\resizebox{0.85\hsize}{!}{\includegraphics{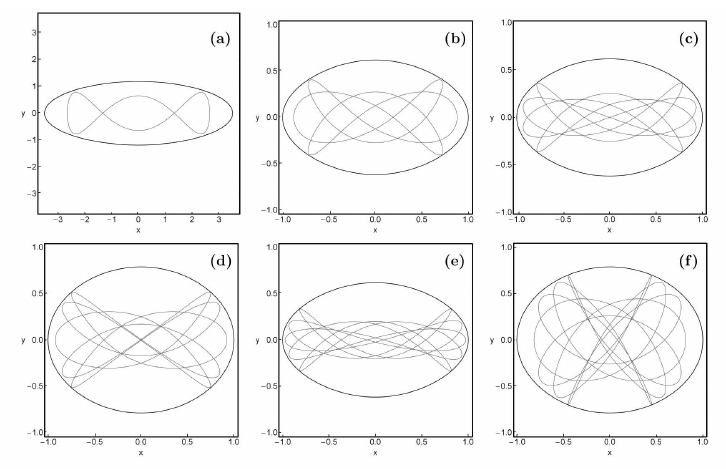}
}

\caption{\baselineskip 3.6mm Characteristic examples of six
secondary resonant orbits in our barred galaxy model: (a) 1:3
resonant orbit; (b) 3:5 resonant orbit; (c) 4:7 resonant type; (d)
5:7 resonant type; (e) 5:9 resonant orbit; (f) 7:9 resonant
orbit.} \label{orbsS}
\end{figure*}

\section{Numerical results}
\label{numres}

In this {s}ection, we shall present all the numerical results of our
research. We numerically integrate several sets of orbits in order
to distinguish between regular and chaotic motion. We use the
initial conditions of orbits mentioned in Section~\ref{compmeth}
in order to construct the corresponding grids, always{ taking}
values inside the Zero Velocity Curve (ZVC) defined by
\begin{equation}
\frac{1}{2} \dot{x}^2 + \Phi_{\rm eff}(x,y=0) = E_{\rm J}.
\label{zvc}
\end{equation}

In most cases, the value of the Jacobi integral was set to $E_{\rm
J} = 1.5$ and kept constant. However in the last subsection where
we investigate the influence of the orbital energy{,} the value of
the Jacobi integral is variable. We chose 
an energy level which gives $x_{\max} \simeq 1$~kpc, where
$x_{\max}$ is the maximum possible value of the coordinate $x$ on
the $(x,\dot{x})$ phase plane, since our study is focused on local
motion of stars. Once the values of the parameters are chosen, we
compute a set of initial conditions as described in
Section~\ref{compmeth} and integrate the corresponding orbits
calculating the value of SALI and then classifying the regular
orbits into different families. Each grid contains roughly a total
of {15\,000} initial conditions $(x_0,\dot{x_0})$ of orbits with
$y_0 = 0$, while $\dot{y_0}$ is always obtained from the Jacobi
integral (Eq.~(\ref{ham})). In each case, we only{ let} one
parameter vary, while all the others have values according to {the
SM} described in Section~\ref{galmod}.

\begin{table*}

\vs \centering

\begin{minipage}{140mm}

\centering
   \caption{ \baselineskip 3.6mm
Type, model and initial conditions of the orbits shown in
Figs.~\ref{orbsM}(a)--(f)  and \ref{orbsS}(a)--(f). In all cases,
$y_0 = 0$ and $\dot{y_0}$ is found from the Jacobi integral, while
$T_{\rm per}$ is the period of the resonant parent periodic
orbits.
   \label{table1}}\end{minipage}

\renewcommand\baselinestretch{1.3}
  \fns\tabcolsep 3.5mm
   \begin{tabular}
   {lcccccccc}
      \hline
      Figure & Type & $\alpha$ & $c_{\rm n}$ & $\Omega_{\rm b}$
       & $E_{\rm J}$ & $x_0$ & $\dot{x_0}$ & $T_{\rm per}$  \\
      \hline
      \ref{orbsM}a &  box          & 4 & 0.50 & 1.0 & 1.5 & --0.91000000 &  0.00000000 &
                -- \\
      \ref{orbsM}b &  1:1          & 2 & 0.25 & 1.0 & 1.5 &  0.41867323 &  0.00000000 &  0.33573521 \\
      \ref{orbsM}c &  1:2 (type a) & 4 & 0.25 & 1.0 & 1.5 &  0.60531937 &  9.66513936 &  0.44222096 \\
      \ref{orbsM}d &  1:2 (type b) & 4 & 0.25 & 1.0 & 593 &  2.84610131 &  0.00000000 &  0.61819761 \\
      \ref{orbsM}e &  2:3          & 4 & 0.25 & 0.0 & 1.5 &  0.00000000 & 13.08286668 &  0.73057544 \\
      \ref{orbsM}f &  3:4          & 4 & 0.25 & 0.0 & 1.5 &  0.09738430 &  7.70724765 &  0.96330681 \\
      \ref{orbsS}a &  1:3          & 9 & 0.25 & 1.0 & 593 &  2.64862858 &  0.00000000 &  0.61518671 \\
      \ref{orbsS}b &  3:5          & 4 & 0.25 & 0.5 & 1.5 &  0.88078044 &  0.00000000 &  1.21951427 \\
      \ref{orbsS}c &  4:7          & 4 & 0.25 & 0.5 & 1.5 &  0.00000000 & 18.96771378 &  1.69575428 \\
      \ref{orbsS}d &  5:7          & 2 & 0.25 & 1.0 & 1.5 &  0.92613552 &  0.00000000 &  2.19852734 \\
      \ref{orbsS}e &  5:9          & 4 & 0.25 & 0.5 & 1.5 & --0.97466418 &  0.00000000 &  2.16072089 \\
      \ref{orbsS}f &  7:9          & 2 & 0.25 & 1.0 & 1.5 &
      0.81782083 &  0.00000000 &  2.86034493 \\
      \hline
   \end{tabular}
\end{table*}

The numerical calculations show that in our barred galaxy model
there are seven main types of orbits: (i) box orbits; (ii) 1:1
resonant orbits; (iii) 1:2 resonant orbits (type a); (iv) 1:2
resonant orbits (type b); (v) 2:3 resonant orbits; (vi) 3:4
resonant orbits, and (vii) chaotic orbits. Apart from the main
families of orbits{,} however, several secondary resonances are also
present.

In Figure~\ref{orbsM}(a)--(f) 
 we present an example of
each of the six basic types of regular orbits, while
Figure~\ref{orbsS}(a)--(f)  shows characteristic examples of the
secondary resonant orbits. The box orbit shown in
Figure~\ref{orbsM}(a)  was computed until $t = 50$ time units,
while all the parent\footnote{For every orbital family there is a
parent (or mother) periodic orbit, that is, an orbit that
describes a closed figure. Perturbing the initial conditions which
define the exact position of a periodic orbit we generate
quasi-periodic orbits that belong to the same orbital family and
librate around their closed parent periodic orbit.} periodic
orbits were computed until one period {was} completed. The black
thick curve circumscribing each orbit is the limiting curve in the
$(x,y)$ plane defined as $\Phi_{\rm eff}(x,y) = E_{\rm J}$.

In Table~\ref{table1} 
 we provide the type, the
initial conditions and the values of the variable parameters for
all the depicted orbits. In the resonant cases, the initial
conditions and the period $T_{\rm per}$ correspond to the parent
periodic orbits. Here we would like to point out that resonant 1:2
type b orbits and resonant 1:3 orbits are only present {in} galaxy
models with relatively high orbital energy $(E_{\rm J} > 100)$.

\begin{figure*}

\vs
\centering \resizebox{0.75\hsize}{!}{\includegraphics{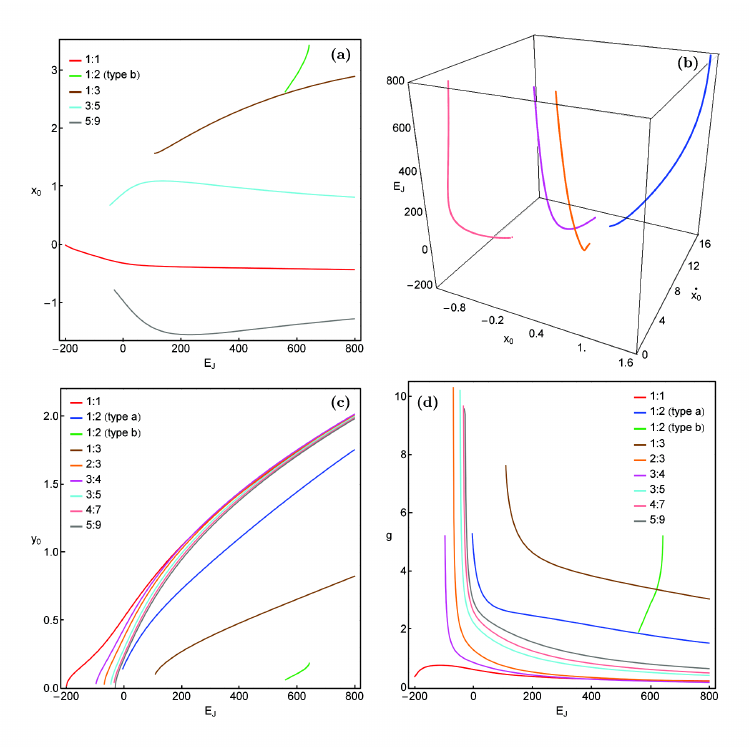}}

\caption{\baselineskip 3.6mm  (a-{\it upper left}): The $(x,E_{\rm
J})$ and (b-{\it upper right}): the $(x,\dot{x},E_{\rm J})$
characteristic curves of the orbital families for SM; (c-{\it
lower left}): The $(y,E_{\rm J})$ characteristic diagram for all
the orbital families; (d-{\it lower right}): A $(g,E_{\rm J})$
diagram showing which types of resonant families support the
galactic bar. The color code is the same in all panels.}
\label{otd}
\end{figure*}

At this point, we would like to clarify some issues regarding the
nomenclature of the orbits in our model. In earlier related
articles{,} orbits in barred galaxies are usually classified
in{to} four main categories: the $x_1$ family which consists of
elongated orbits along the bar, families $x_2$ and $x_3$ the
orbits of which are also elongated but perpendicular to the bar
and the retrograde $x_4$ family \citep[e.g.,][]{CP80}. In the
present case however, we decided to follow for consistency the
same classification used in all previous papers of this series{,}
according to which the orbits are separated into three main
categories: (i) box orbits, (ii) $n:m$ resonant orbits, and (iii)
chaotic orbits. According to our notation, all resonant orbits
have the following recognizable $n:m$ oscillatory pattern: a
resonant orbit completes $m$ oscillations perpendicular to the
major axis of the bar in the time that it takes the orbit to
perform $n$ circuits along the major axis. Furthermore, {an} $n:m$
resonant orbit would be represented by $m$ distinct islands of
invariant curves in the $(x,\dot{x})$ phase plane and $n$ distinct
islands of invariant curves in the $(y,\dot{y})$ surface of
section. In our research, we searched for resonant orbits $n:m$ up
to $n,m \leq 9${;} therefore, for all higher resonant orbits the
numerical code assigns ``box" classification (this is a usual
technique in orbit classification), which is correct for $n \neq
m$ (high resonant box orbits, e.g., \citealt{CB82}). As was pointed
out in the review of \citet{SW93} (p. 31), there are several
different notations regarding the naming of the orbital families
in barred galaxies. Traditionally, orbits are named $m:l$, where
$m$ denotes the number of radial oscillations an orbit performs
before it closes, while $l$ corresponds to the{ number of} turns
of the orbit around the center of the potential. Throughout the
paper, we shall use the first notation which{,} from our point of
view, is more descriptive and better{ fits} our computational
procedures.

It is of particular interest to determine which types of regular
orbits support the barred structure in our galactic model. Looking
carefully{ at} the main types of orbits shown in
Figure~\ref{orbsM}(a)--(f)  it can be seen that the box and the
1:2 (types a and b) resonant orbits are the ones we are looking
for.
We also observe in Figure~\ref{orbsS}(a)--(f) 
 that{,}
apart from the 7:9 resonant family{,} all the other secondary
resonances may support, more or less, the barred structure. In
order to quantify our search for orbits supporting the bar, we
must define a mathematical criterion thus distinguishing which
types of orbits have the ability to support the bar. This issue
can be solved if we exploit the geometry of the orbits. In
particular, we compute the maximum values of the $x$ and $y$
coordinates of the regular orbits, $x_{\max}$ and $y_{\max}$
respectively, along the numerical integration. Then, the ratio $g
= x_{\max}/y_{\max}$ defines whether an orbit supports the barred
structure or not. The threshold value regarding the ratio strongly
depends on the strength of the bar. Our numerical calculations
suggest that a safe threshold for the ratio is the value
$\sqrt{\alpha}$. Therefore, all types of regular orbits with
values $g \geq \sqrt{\alpha}$ can support the barred structure.

One of the most interesting structures that are often observed in
barred galaxies are rings. There are three main types: (i) nuclear
rings situated near the central nucleus, (ii) inner rings
surrounding the bar and (iii) outer rings with a relatively extended
diameter. In this work, we investigate the orbital properties near
the central region of the barred galaxy and therefore, we {only }focus our
study {on} nuclear rings. Bars have a natural
tendency to concentrate gas near the nucleus but also can setup
resonances which usually act as focal points for the gas flow. The
reader can find more details about rings in barred galaxies in the
review of \citet{K04}. We argue that 1:1 resonant orbits with
approximately $0.5 < g < 1.5$ (both prograde and retrograde)
{are} the best candidate{s for} supporting the nuclear ring
structure in barred galaxies. Thus, we shall pay special attention
{to} how the basic parameters of the system affect the
amount of 1:1 resonant orbits.

In Figure~\ref{otd}(a), 
 we present a very informative
diagram{,} the so-called ``characteristic" orbital diagram
\citep{CM77,CB85,CM85} for SM (except the 1:3 resonance for which
$\alpha = 9$). It shows the evolution of the $x$ coordinate of the
initial conditions of the parent periodic orbits of each orbital
family as a function of their orbital  energy $E_{\rm J}$ (Jacobi
constant). Here we should emphasize that for orbits starting
perpendicular to the $x$-axis, we{ only} need the initial
condition of $x_0$ in order to locate them on the characteristic
diagram. On the other hand, for orbits not starting perpendicular
to the $x$-axis (i.e., the 1:2 type a, 2:3, 3:4 and 4:7 families)
initial conditions {like} position-velocity pairs $(x,\dot{x})$
are required and{,} therefore, the characteristic diagram is now
{3D,} providing full information regarding the interrelations of
the initial conditions in a tree of families of periodic orbits
(Fig.~\ref{otd}(b)). Furthermore, the diagram shown in
Figure~\ref{otd}(c) is another type of ``characteristic" diagram
\citep{SW93,BT08,Z13}, where the value of the Jacobi integral
$E_J$ is plotted against the coordinate {where} the minor axis of
the bar{ crosses the} $y${-axis}. As {can be} seen in
Figures~\ref{orbsM} and \ref{orbsS}{,} all the higher resonant
orbits encountered in our potential (i.e., the 2:3, 3:4, 3:5, 4:7
and 5:9 resonant orbits) have complicated shapes thus crossing{
the $y$-axis} multiple times and {at} several positions. When
constructing the diagram shown in Figure~\ref{otd}(c), we
considered {where} all these higher resonant orbits cross{ed} with
higher absolute value{s} of $y_0$.

In the same vein, we decided to create a new type of diagram which
is called the ``support diagram" and it is presented in
Figure~\ref{otd}(d).  In this diagram, we see the evolution of the
$g$ parameter of the parent periodic orbits {for} each family as a
function of the energy $E_{\rm J}$ for SM. The aim of this plot is
to help us decide which types of resonant orbits support the bar
and which do not. The diagram works as follows: the higher a
curve{ is} (in other words, greater values of $g$) corresponding
to a particular resonant family{,} the more supportive the barred
structure is {for }this orbital family. We observe that both types
of the 1:2 family as well as the 1:3 family highly support the
bar, while on the other hand the 1:1 family has no contribution
whatsoever to the bar (this family favors nuclear ring formation).
Here we should note that the 1:2 type b family bifurcates from the
main 1:2 type a family and it is {only }present at relatively high
energy levels $(561 < E_{\rm J} < 642)$. Moreover, we may say that
in general terms higher resonant orbits such as the 3:5, 4:7 and
5:9 families can also support the bar, {but} the 2:3 and 3:4
resonant families can {only }support{ it} at low energies.

\begin{figure}
\centering
\includegraphics[width=0.78\hsize]{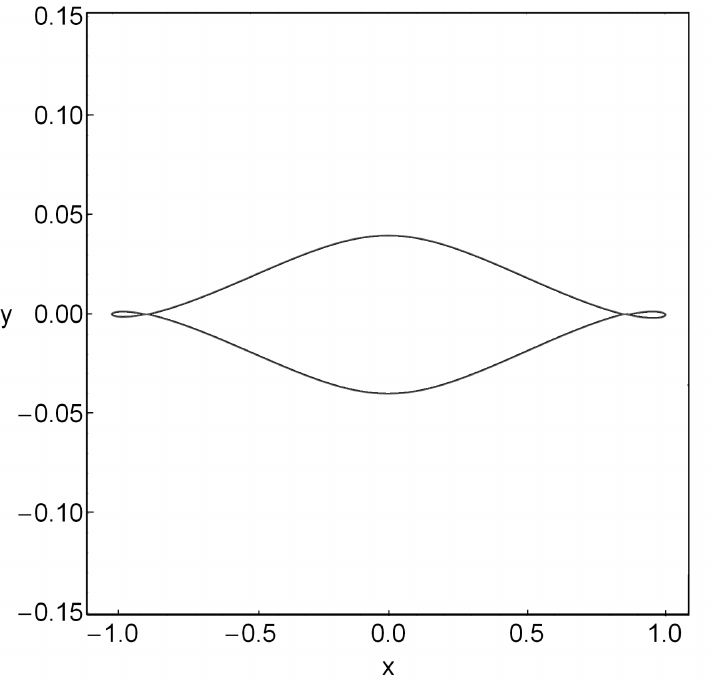}

\caption{\baselineskip 3.6mm A highly unstable $x_1$ periodic
orbit in SM. The initial conditions and more details are given in
the text.} \label{orbx1}
\end{figure}

It is widely accepted that galactic bars are in fact created by
regular orbits which circulate around the so-called ``$x_1$"
periodic orbits (e.g., Skokos et al.\ 2002a,b)
The $x_1$ orbits are elongated along the
bar's major axis and usually have the shape of simple ellipses.
However, with increasing energy, they can form a cusp and even two
loops at the extremities.

In Figure~\ref{orbx1}, 
 we present a highly unstable
(Stability Index (S.I.) $= -17.24$) $x_1$ periodic orbit in SM
with initial conditions: $x_0 = 0.86508078$, $y_0 = 0$, $\dot{x_0}
= 6.31880018$, while the value of $\dot{y_0}$ was obtained from
the Jacobi integral. Our numerical calculations indicate that{,}
over the entire range of $E_{\rm J}$, our galactic model does not
support the $x_1$ orbital family. This is true because the
position of the parent periodic orbit in the $(x,\dot{x})$ phase
plane is deeply buried in the chaotic region (see
Fig.~\ref{gridsa}(c)) 
 without any indication of an
existing stability island around the periodic point which could
support $x_1$ quasi-periodic orbits. Moreover, our computations
reveal that in our barred galaxy model the $x_1$ periodic orbits
are so unstable ($\rm |S.I.| > 10$) that it is impossible to
create the corresponding characteristic curve (see
Fig.~\ref{otd}).

In the literature, there are some isolated examples of barred galaxy
models, such as the ``Cazes" bar \citep{BT01}, the ``model B" of
\citet{SPA02b} and the ``propeller" orbits in \citet{KP05}{,} all{ of}
which{ have} the $x_1$ orbital family{ that} is no longer the dominant
one and other types of regular orbits play the leading role in
supporting the barred structure. Thus, our simple galactic model can
be considered as a member of this closed group. In our paper, we
demonstrate that the role of $x_1$ orbits can be successfully
supplemented by other types of regular families of orbits which can
also support the bar. From a mathematical point of view, whether an
orbital family is present or not strongly depends on the choice of
the potential describing the properties of the bar. The vast
majority of papers devoted {to} the subject of
barred galaxies utilize a Ferrers potential \citep{F77} in
order to model the bar. {As it} happens, the traditional
$x_1$ orbital family dominate{s} in galaxy models where
a Ferrers potential is used. However, this does not
necessarily mean that $x_1$ orbits should {always }dominate
the structure of the phase plane of all {2D}
bar potentials used to model realistic galactic bars. In fact, in
the present paper, we present evidence that numerous resonant orbits
can also support the barred structure of a galaxy{ equally well}.

\subsection{Influence of the Strength of the Bar}
\label{str}

To explore the influence of the strength of the bar $\alpha$ on
the orbital structure of the barred galaxy, we let it vary while
fixing the numerical values of all the other parameters in our
model according to SM and integrate orbits for the set $\alpha =
\{1,2,3, ..., 10\}$. Once the values of the parameters were
chosen, we computed a set of initial conditions as described in
Section~\ref{compmeth} and integrated the corresponding orbits{,}
computing the SALI of the orbits and then classifying the regular
orbits into different families. Here we should point out that
when $1.1 < \alpha < 2$ our model describes the properties of a
weak rotating bar, {but} when $\alpha \geq 2$, we have the
presence of a strong bar.

\begin{figure*}

\vs \centering
\resizebox{0.82\hsize}{!}{\includegraphics{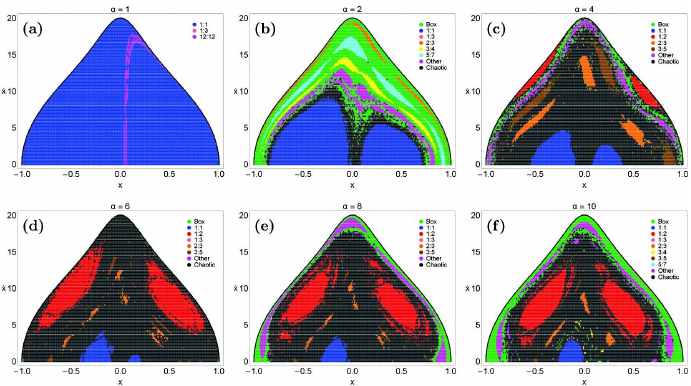}}

\vs

\caption{\baselineskip 3.6mm Orbital structure of the
$(x,\dot{x})$ phase plane of the barred galaxy model for different
values of the strength of the bar $\alpha$.} \label{gridsa}

\vs
\vs \centering
\resizebox{0.8\hsize}{!}{\includegraphics{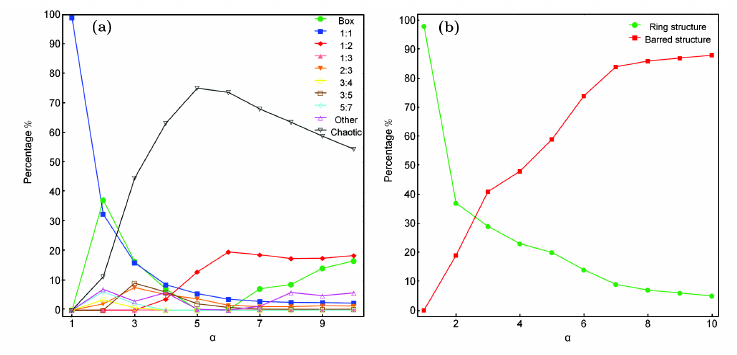}}

\caption{\baselineskip 3.6mm Evolution of the percentages of
(a-{\it left}): the different types of orbits in our barred galaxy
model and (b-{\it right}): the types of regular orbits supporting
the formation of nuclear rings {or} {a }barred structure, when
varying the strength of the bar $\alpha$.} \label{percsa}
\end{figure*}

In Figure~\ref{gridsa}(a)--(f)  we present six grids of initial
conditions $(x_0,\dot{x_0})$ of orbits that we have classified for
different values of the strength of the bar $\alpha$. The
numerical calculations reveal that when $\alpha$ varies there are
eight main types of regular orbits. All the different regular
families can be identified by the corresponding sets of islands
which are formed in the phase plane. In particular, we see the
eight main families already mentioned: (i) box orbits occupying
the outer parts of the phase plane; (ii) 1:1 resonant orbits
surrounding the two central main periodic point; (iii) 1:2
resonant orbits (type a) producing two stability islands; (iv) 1:3
resonant orbits generating three tiny islands at the outer parts
of the grid; (v) 2:3 resonant orbits displaying a set of three
islands; (vi) 3:4 resonant orbits forming a set of four islands;
(vii) 3:5 resonant orbits producing a chain of five islands at the
outer parts of the phase plane and (viii) 5:7 resonant orbits
corresponding to a set of seven islands inside the box region. The
term ``other" refers to all different types of resonant orbits
with $n,m < 9$ not included in the former categories. It is seen
that apart from the several regions of regular motion, we observe
the presence of a unified chaotic sea which {surrounds} all the
islands of stability. The outermost black thick curve is the ZVC
defined by Equation~(\ref{zvc}).

When $\alpha = 1$ the bar does not exist, the total potential is
integrable and as expected there is no evidence whatsoever of
chaotic motion in the phase plane of Figure~\ref{gridsa}(a).
 Almost the entire grid is covered by initial
conditions corresponding to 1:1 resonant orbits, while we observe
a thin layer of higher resonant 12:12 orbits (which is a
bifurcation of the main 1:1 resonant family) at the prograde $(x >
0)$ side of the phase plane. The structure of the phase plane{,}
however, changes drastically when $\alpha = 2$. We observe in
Figure~\ref{gridsa}(b) that the area occupied by 1:1 orbits has
been reduced and there are two distinct regions of stability; one
at the right part of the grid corresponding to prograde 1:1 orbits
and one at the left side corresponding to retrograde 1:1 orbits.
We see{,} on the other hand, that box orbits swarm the outer parts
of the grid, while several resonant families such as the{ cases
of} 2:3, 3:4, 4:5, 5:7 and 7:9 emerge inside the box area, thus
producing sets of multiple stability islands. We also have to
notice the presence of a small chaotic layer which
 {defines the separation} between 1:1 and box orbits. Things
change even more as the strength of{ the} bar increases.
Figure~\ref{gridsa}(c) shows the structure of the phase plane when
$\alpha = 4${,} which corresponds to SM. It is evident that the
chaotic layer has been transformed into a vast chaotic sea
flooding the majority of the phase plane. The amount of box and
higher resonant orbits has been reduced significantly and those
orbits are confined to the outer parts of the grid. Two additional
resonances{,} that is the 1:2 (type a) and the 3:5{,} emerge. When
$\alpha = 4$ the potential of the bar is at integer resonance
{1:$\sqrt{\alpha}$=1:2}{,} so we anticipated the existence of the
1:2 resonance. In Figure~\ref{gridsa}(d) where $\alpha = 6$, we
observe that the extent of the chaotic sea has grown even
further{,} mainly at the expense of box and 1:1 orbits. In fact,
there are only {a }few isolated points in the grid corresponding
to box orbits. On the contrary, the 1:2 stability islands have
{more than }doubled their size.

Figure~\ref{gridsa}(e) and (f){,} where $\alpha = 8 $ and $\alpha
= 10$ respectively{,} indicate{s} that a further increase {in} the
strength of the bar has only {a }minor {influence} {on }the
orbital structure of the phase plane. This is true because the
amount of chaotic and 1:2 (type a) orbits seems to saturate. The
most visible differences are the following: (i) the 1:1 prograde
stability island disappears; (ii) box orbits gain ground again and
(iii) some secondary resonance{s} such as the 3:7 {and} 2:5 appear
inside the box region. It should be noticed that the islands
{representing} the 1:3 resonance are so small that they appear as
{isolated} points in the grids of Figure~\ref{gridsa}(a)--(f).

Figure~\ref{percsa}(a) 
 shows the resulting percentages
of chaotic orbits and {those} of the main families of regular
orbits as $\alpha$ varies. It can be seen that there is a strong
correlation between the percentages of most types of orbits and
the value of the strength of the bar. When the bar is absent
$(\alpha = 1)$, the entire phase plane is covered by 1:1 resonant
orbits. As the strength of the bar is increased{,} however, the
percentage of 1:1 resonant orbits decreases rapidly {at} an
exponential rate. At the same time, the percentage of chaotic
orbits increases and when $\alpha > 3$ chaotic orbits {are} the
most populated family, always occupying  more than 50\% of the
phase plane. In particular, the largest amount of chaos, around
75\%, is observed when $\alpha = 5$. On the other hand, when
$\alpha > 5$ the percentage of chaos is reduced almost linearly.
The box orbits exhibit the peak of their percentage (around 40\%)
at $\alpha = 2$ and then for $2 < \alpha \leq 6$ their rate is
reduced, while for $\alpha > 6$ this tendency is reversed. The
percentage of the 1:2 resonant orbits starts to grow when $\alpha
> 3$ and it seems to saturate around 20\% when $\alpha > 6$. At
higher value{s} of the strength of the bar{ that are} studied, the
percentages of box and 1:2 resonant orbits (type a) tend to a
common value (around 20\%), thus sharing two{-}fifths of the
entire phase plane. It is evident that in barred galaxies, varying
the strength of the bar mainly shuffles the orbital content of all
the other resonant orbits, whose percentages present fluctuations
at low values (less than 10\%). Thus, taking into account all the
above, we could say that in barred galaxy models the strength of
the bar $\alpha$ {mostly }influences box, 1:1, 1:2 (type a) and
chaotic orbits. In fact, a large portion of 1:1 and box orbits
turn into 1:2 (type a) and chaotic orbits as the bar becomes
stronger, or in other words, as the value of $\alpha$ increases.

The evolution of the percentages of regular orbits supporting a
ring or barred structure as a function of the strength of the bar
$\alpha$ is shown in Figure~\ref{percsa}(b). {As w}e explained
previously, we assume that only the 1:1 resonant orbits with $0.5
< g < 1.5$ support the formation of nuclear rings, while all types
of regular orbits with $g \geq \sqrt{\alpha}$ support the barred
structure. Here we have to point out that the percentages do not
refer to the total number of tested orbits (regular plus chaotic)
in each grid but{ rather} to the total regular orbits. It is seen
that for small values of $\alpha$ $(\alpha < 2.5)$, {as} is the
case of a weak bar, almost all the regular orbits favor the
formation of nuclear rings. This is true because{,} as was
discussed previously in Figure~\ref{gridsa}(a)--(b){,} the
majority of the phase plane is covered by 1:1 regular orbits.
However, as the value of $\alpha$ increases and the bar gains
strength{,} the rate of 1:1 orbits which support rings decreases
rapidly, while at the same time the percentage of regular orbits
supporting {a }barred structure grows steadily and for $\alpha >
7$ it seems to saturate around 85\%. Therefore, we may conclude
that rings are highly favored in weak bar models, {but} in galaxy
models possessing strong bars only roughly 5\% of orbits support
nuclear rings.

\subsection{Influence of the Scale Length of the Nucleus}
\label{nsl}

Our next step is to reveal how the overall orbital structure in our barred galaxy model is affected by the scale length of the nucleus $c_{\rm n}$. As usual, we let this quantity vary while fixing the values of all the other parameters according to SM and integrating orbits in the phase plane for the set $c_{\rm n} = \{0.05,0.10,0.15,...,0.50\}$. Our numerical calculations show that most of the main regular families are the same as in the previous case. The only difference is that in the group of regular families the 1:3 and 5:7 resonant families are now substituted by the 4:7 and 5:8 resonant families.

\begin{figure*}
\centering
\resizebox{0.82\hsize}{!}{\includegraphics{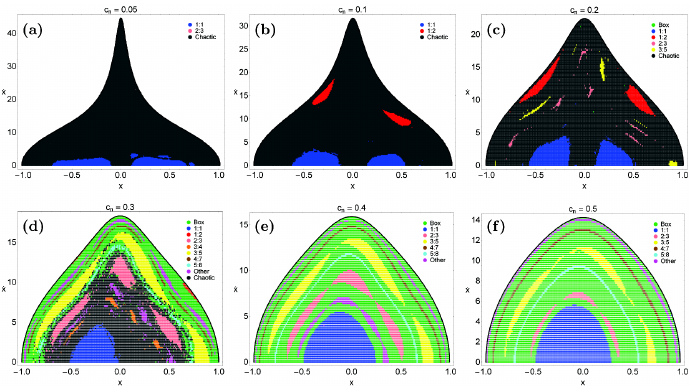}} 

\caption{\baselineskip 3.6mm Orbital structure of the
$(x,\dot{x})$ phase plane of the barred galaxy model for different
values of the scale length $c_{\rm n}$ of the central spherical
nucleus. \label{gridscn}}

\vs \centering
\resizebox{0.82\hsize}{!}{\includegraphics{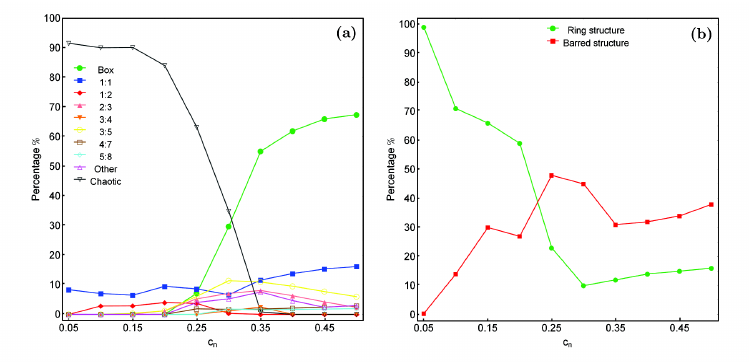}}

\caption{\baselineskip 3.6mm Evolution of the percentages of
(a-{\it left}): the different types of orbits in our barred galaxy
model and (b-{\it right}): the types of regular orbits supporting
the formation of nuclear rings {or a} barred structure, when
varying the scale length $c_{\rm n}$ of the central spherical
nucleus.} \label{percscn}
\end{figure*}

Figure~\ref{gridscn}(a)--(f) 
 depicts six grids of
initial conditions $(x_0,\dot{x_0})$ of orbits that we have
classified for different values of the scale length of the nucleus
$c_{\rm n}$. Again, all the different regular families can be
identified by the corresponding sets of islands which are formed
in the phase plane. It is observed in Figure~\ref{gridscn}(a) that
when the central nucleus is very dense, that is when $c_{\rm n} =
0.05$, the vast majority of the phase plane is covered by chaotic
orbits, {but} only the 1:1 resonant family survives. There is also
weak evidence of the 2:3 resonance{,} however, the corresponding
initial conditions are few and deeply buried in the vast chaotic
sea. As we increase the value of $c_{\rm n}$ and consequently the
central nucleus becomes less dense{,} we see in
Figure~\ref{gridscn}(b) ($c_{\rm n} = 0.1$) and (c) ($c_{\rm n} =
0.2$) that additional regular families such as box, 1:2 (type a)
and 3:5 emerge inside the chaotic sea. Moreover, the extent of the
1:1 and 1:2 stability islands grows as the scale length of the
nucleus increases. The structure of the phase plane becomes very
interesting in Figure~\ref{gridscn}(d) where $c_{\rm n} = 0.3$.
This is true for many reasons. First of all, the amount of chaos
is decreased and the prograde 1:1 stability island disappears. At
the outer parts of the phase plane, box orbits take control, {but}
the 1:2 stability islands are significantly reduced. It can also
be seen that the entire phase plane is swarmed by many types of
resonant orbits producing several sets of stability islands. In
particular, the 2:3 and 3:5 are the most populated resonant
families.

In Figure~\ref{gridscn}(e) and (f) we present the cases where
$c_{\rm n} = 0.4$ and $c_{\rm n} = 0.5$ respectively, that is when
the central nucleus is sparse enough. In both cases, the structure
of the grid is similar and we observe that the entire phase plane
is {only }covered by regular orbits. In fact, a large portion of
the grid corresponds to box orbits, {but} all the resonant
families exist inside the box region. We should notice that the
1:2 resonance is completely absent when the scale length of the
spherical nucleus obtains high values. Furthermore, the 1:1
stability island is now located almost at the center of the grid,
thus containing a mixture of prograde and retrograde 1:1 orbits.
It should also be pointed out that as the central nucleus becomes
more dense (small values of $c_{\rm n}$) there is an {increase} in
the allowed velocity $\dot{x}$ of the stars near the center of the
galaxy.

The resulting percentages of chaotic and regular orbits for the
barred galaxy model as the scale length of the nucleus $c_{\rm n}$
varies are shown in Figure~\ref{percscn}(a). 
 It is
evident that box and chaotic orbits are the types of orbits mainly
affected by the scale length of the nucleus. In particular, we see
that when the central nucleus is very dense ($c_{\rm n} < 0.15$)
the motion of stars is highly chaotic since about 90\% of the
entire phase plane is covered by chaotic orbits. However, as the
values of $c_{\rm n}$ increase and the nucleus becomes less dense
($c_{\rm n} > 0.2$){,} the percentage of chaos displays a sharp
decrease and eventually vanishes when $c_{\rm n} > 0.35$. The rate
of box orbits{,} on the other hand, {only }start{s} to grow when
the nucleus is sparse enough ($c_{\rm n} > 0.2$) and when $c_{\rm
n} > 0.3$ box orbits {are} the dominant type occupying about
two{-}thirds of the phase plane. It {can be} seen in
Figure~\ref{percscn}(a) that all the other types of resonant
orbits are considerably less affected by the shifting of $c_{\rm
n}$. The rate of 1:1 resonant orbits exhibit{s} minor fluctuations
for small values of the scale length{,} however, when $c_{\rm n}
> 0.3$ it increases, while at the same time the percentage of the
1:2 (type a) resonant orbits vanishes. The evolution of the
percentages of the 2:3, 3:5 and higher resonant families of orbits
exhibit a similar pattern; they start to grow when $c_{\rm n} >
0.15$ and they decrease for $c_{\rm n} > 0.35$. The percentages of
the 4:7 and 5:8 resonant families on the other hand, {only }have non-zero
values when $c_{\rm n} > 0.25$ and then it seems they saturate
around 2\%. Therefore, increasing the scale length of the nucleus
(in other words the nucleus becomes less concentrated and dense) in
barred galaxy models turns mainly chaotic orbits into box orbits,
while resonant orbits are less affected.

Figure~\ref{percscn}(b) shows the evolution of the percentages of
regular orbits supporting{ a} ring {or} barred
structure\footnote{According to SM $\alpha = 4$, thus the
threshold value for the ratio $g$ is $\sqrt{\alpha} = 2$.} as a
function of the scale length of the nucleus $c_{\rm n}$. We
observe that when the central nucleus is dense enough the vast
majority of regular orbits support nuclear rings{,} however, as
the nucleus becomes less dense the rate of regular orbits that
support the bar grows constantly and when $c_{\rm n} > 0.25$ it
{dominates}. At the same time, the percentage of ring structure
orbits decreases rapidly and saturates around 20\% when $c_{\rm n}
> 0.3$. Thus, one may conclude that in barred galaxy models with
dense nuclei only the formation of nuclear rings is favored, {but}
when the central nucleus is sparse enough about 40\% of the total
types of regular orbits support the barred structure and only 20\%
of them {support} nuclear rings.

\subsection{Influence of the Angular Velocity}
\label{angv}

The next parameter under investigation is the angular velocity
$\Omega_{\rm b}$ of the bar. We shall try to understand how the
overall orbital structure in our barred galaxy model is influenced
by this parameter. Again, we let this quantity vary while fixing the
values of all the other parameters of our galactic model according
{to }SM and integrating orbits in the phase plane for the set
$\Omega_{\rm b} = \{0,0.25,0.5,...,2.5\}$. The numerical experiments
indicate that in this case, the main families of regular orbits are
similar to those discussed in subsection \ref{nsl}.

\begin{figure*}

\vs \centering
\resizebox{0.8\hsize}{!}{\includegraphics{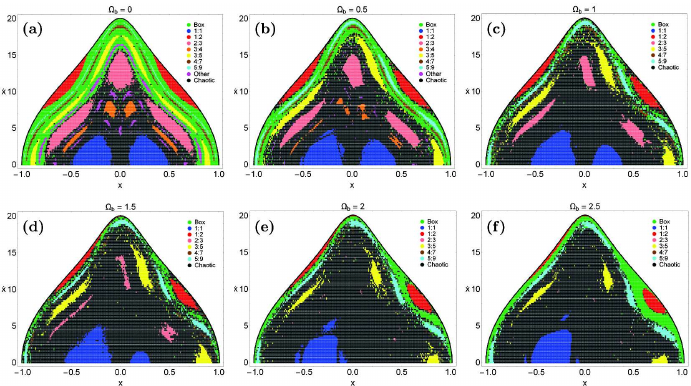}
}

\caption{ \baselineskip 3.6mm Orbital structure of the
$(x,\dot{x})$ phase plane of the barred galaxy model for different
values of the angular velocity $\Omega_{\rm b}$ of the bar.}
\label{gridsR}
\centering


\vs \resizebox{0.78\hsize}{!}{\includegraphics{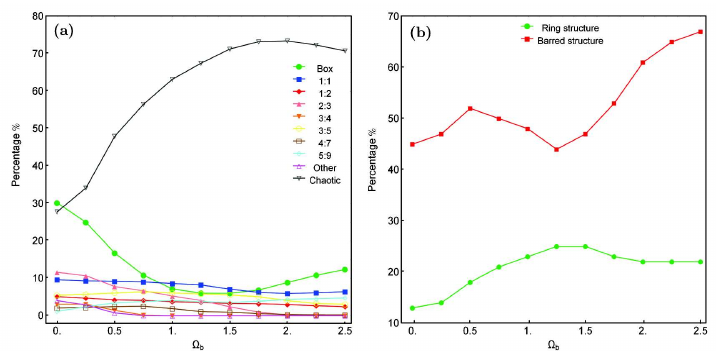}}

\caption{\baselineskip 3.6mm  Evolution of the percentages of
(a-{\it left}): the different types of orbits in our barred galaxy
model and (b-{\it right}): the types of regular orbits supporting
the formation of nuclear rings {or a} barred structure, when
varying the angular velocity of the bar $\Omega_{\rm b}$.}
\label{percsR}
\end{figure*}

In order to explore the structure of the phase plane when
$\Omega_{\rm b}$ varies, we integrated orbits in several grids. A
sample of six grids of initial conditions $(x_0,\dot{x_0})$ of
orbits that we have classified for different values of the angular
velocity is given in Figure~\ref{gridsR}(a)--(f). 
 As
usual, all the different types of regular families can{ be} easily
identified by the corresponding sets of islands which are produced
in the phase plane. In Figure~\ref{gridsR}(a) we present the case
where $\Omega_{\rm b} = 0$ which means that the bar does not
rotate. Due to the absence of rotation{,} the Coriolis force is
zero and therefore, the phase plane is symmetrical{ with respect}
to the $x = 0$ axis. We observe a unified chaotic domain at the
central parts of the phase plane, {but} there are also numerous
stability islands corresponding to resonant families. When
$\Omega_{\rm b} = 0.5$, {which} is a model of a slow{ly}
rotat{ing} bar, it can be seen in Figure~\ref{gridsR}(b) that the
structure of the phase displays minor differences with respect to
Figure~\ref{gridsR}(a), {with} the growth of the region occupied
by chaotic orbits and the depopulation of box, 2:3, 3:4 and higher
resonant orbits{ being most visible}. As the bar gains speed, this
tendency is continued in Figure~\ref{gridsR}(c) ($\Omega_{\rm b} =
1$) and (d) ($\Omega_{\rm b} = 1.5$) and as a result the 3:4 and
higher resonant orbits are completely absent. In
Figure~\ref{gridsR}e where $\Omega_{\rm b} = 2$, we see that the
prograde 1:1 stability island disappears, while the 2:3 resonance
has been depopulated so much that the corresponding initial
conditions appear as {isolated} points in the grid. The region of
box orbits, on the other hand, seems to increase. It is evident
from Figure~\ref{gridsR}(f), that in the case of a {fast}
rotat{ing} bar ($\Omega_{\rm b} = 2.5$){,} the region of box
orbits increases even further{,} thus suppressing the 1:2 (type a)
stability islands. One may reasonably conclude that the faster the
rotation of the bar{ is}, the more chaos is observed in the barred
galaxy.

Figure~\ref{percsR}(a) 
 shows the evolution of the
resulting percentages of chaotic and regular orbits for the barred
galaxy model as the angular velocity $\Omega_{\rm b}$ varies. Once
more we see that the angular velocity mainly affects the
{fraction} of box and chaotic orbits. When there is no rotation of
the bar ($\Omega_{\rm b} = 0$), box and chaotic orbits seem to
share about 60\% of the phase plane. As the value of the angular
velocity increases and the bar gains speed{,} we observe that the
percentages of both box and chaotic orbits evolve similarly but
with different direction{s}. Being more precise, the rate of box
orbits decreases until $\Omega_{\rm b} = 1.25$, {but} for larger
values of the angular velocity {it} exhibits an increase. The
percentage of chaotic orbits, on the other hand, increases
rapidly{,} however, when $\Omega_{\rm b} > 1.75$ it displays a
minor decrease. Nevertheless, the motion of stars is highly
chaotic throughout as the percentage of chaotic orbits remains
larger than any other individual regular family. In fact, in most
cases more than half of the phase plane is occupied by chaotic
orbits and the peak (around 75\%) is observed when $\Omega_{\rm b}
= 1.75$. One may see in Figure~\ref{percsR}(a) that all the other
types of resonant orbits are considerably less affected by the
shifting of $\Omega_{\rm b}$. The rate of the 1:1 resonant orbits
decreases{,} especially for $\Omega_{\rm b} > 1.25$ when the
prograde 1:1 stability islands starts to shrink until it
disappears completely from the phase plane. In the same vein, the
percentage of the 1:2 resonant (type a) family exhibits a minor
almost linear decrease. The rate of the 2:3 family{,} on the
contrary, shows a rapid reduction and vanishe{s} when $\Omega_{\rm
b} > 2$. The 3:4 and higher resonant orbits seem to be unable to
cope with the rotation of the bar and their rates are zeroed very
quickly even at low speed ($\Omega_{\rm b} > 0.75$). Here we
should point out that only the 5:9 resonant family is favored by
the rotation of the bar since it is the only regular family {that}
constantly{ augments} its rate with increasing $\Omega_{\rm b}$.
Therefore, increasing the angular velocity of the bar general{ly}
turns different types of regular orbits into chaotic{ ones}.

The evolution of the percentages of regular orbits supporting
either{ a} ring or barred structure as a function of the angular
velocity of the bar $\Omega_{\rm b}$ is given in
Figure~\ref{percsR}(b). We observe that the majority of regular
orbits support the barred structure throughout, {but} only about
one{-}fifth of the total regular orbits support the formation of
nuclear rings. In particular, for relatively small speeds
($\Omega_{\rm b} < 1.25$) the rate of ring structure orbits
increases{,} reaching about 25\%, while for larger values of the
angular velocity it saturates around 20\% of the total regula{r}
orbits. At the same time, the rate of regular orbits supporting {a
}barred structure displays a small fluctuation around 45\% and
only in models with fast rotating  bars
($\Omega_{\rm b} > 1.25$) 
increases rapidly{,} occupying more than two{-}thirds of regular
orbits. Therefore, we may conclude that slowly rotat{ing} bars
support the formation of nuclear rings, {but} fast rotat{ing}
bars{ mainly} favor the barred structure.

\subsection{Influence of the Energy}
\label{ener}

The last parameter under investigation is the total orbital energy
$E_{\rm J}$. In order to explore how the energy level affects the
overall orbital structure of our barred galaxy model, we use the
normal procedure according to which we let the energy vary while
fixing the values of all the other parameters of our galactic
models according to SM. At this point{,} we should point out that
the particular value of the energy determines the maximum possible
value of the $x$ coordinate $(x_{\max})$ on the $(x,\dot{x})$
phase plane. To select the energy levels, we chose those values of
the energy which give $x_{\max} = \{0.5,1,1.5,...,4\}$. Our
numerical computations show that the main families of regular
orbits remain the same as those discussed in the previous two
subsections.

\begin{figure*}[!tH]

\centering
\resizebox{0.85\hsize}{!}{\includegraphics
{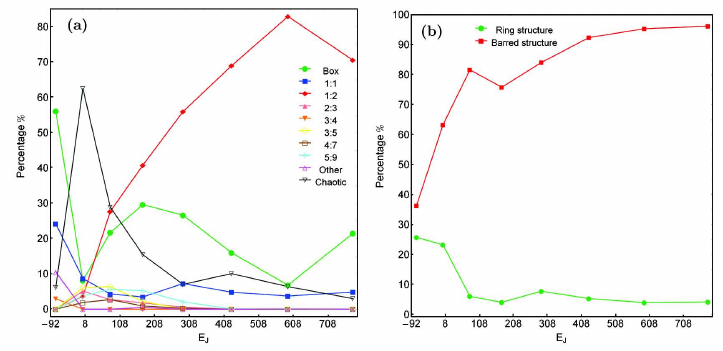}}

\caption{\baselineskip 3.6mm Orbital structure of the
$(x,\dot{x})$ phase plane of the barred galaxy model for different
values of the Jacobi integral $E_{\rm J}$.} \label{gridsE}

\centering

\vs\vs

\includegraphics[width=0.46\hsize]{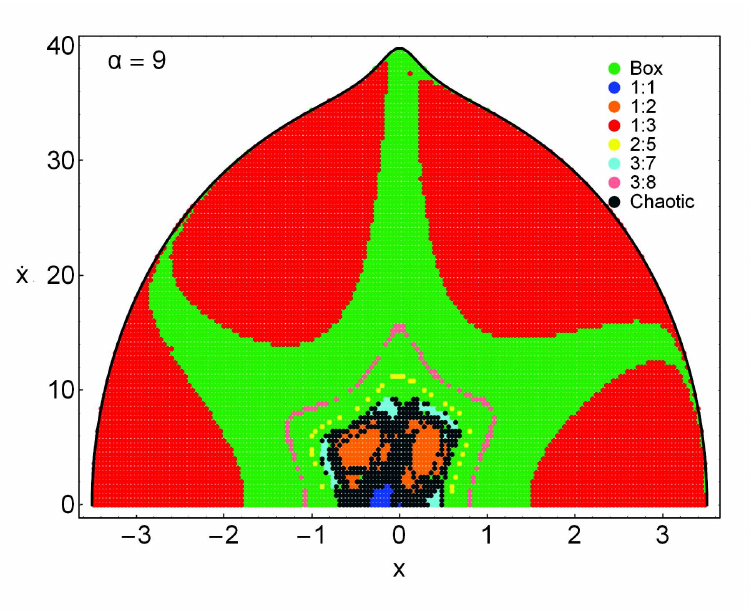}

\caption{\baselineskip 3.6mm Orbital structure of the $(x,\dot{x})$
phase plane of the barred galaxy model when $\alpha = 9$ and $E_{\rm
J} = 593$, while the values of all the other parameters are{ defined}
according to SM. We observe that the 1:3 resonant orbits
occupy the vast majority of the phase plane.} \label{R13}
\end{figure*}

Figure~\ref{gridsE}(a)--(f) 
 shows six grids of initial
conditions $(x_0,\dot{x_0})$ of orbits that we have classified for
different values of the Jacobi integral $E_{\rm J}$. In
Figure~\ref{gridsE}(a){,} $E_{\rm J} = -77$ which corresponds to
local motion of stars moving very close to the central nucleus
with $x_{\max} =  0.5$ kpc. It is seen that the vast majority of
the phase plane is covered by initial conditions of regular
orbits, while a weak chaotic layer exists at the central parts of
the grid separating regions of box and 1:1 resonant orbits. We
also observe the presence of several chains of stability islands
inside the box region. These sets of stability islands are
produced by secondary resonances such as the 4:5, 5:6, 5:7, 6:7
and 7:9 families. The grid shown in Figure~\ref{gridsE}(b) where
$E_{\rm J} = 1$ is very similar to those discussed earlier in
Figures~\ref{gridsa}(c) and \ref{gridsR}(c).
Figure~\ref{gridsE}(c) shows a grid where $E_{\rm J} = 80$ and
$x_{\max} = 1.5$~kpc. Here the area occupied by chaotic orbits is
reduced, as there is a considerable increase of the 1:2 (type a)
stability islands. The increase {in} the amount of 1:2 (type a)
resonant orbits continues in Figure~\ref{gridsE}(d) where $E_{\rm
J} = 290$. In this case, box and 1:2 resonant orbits (type a)
share the majority of the phase plane. A weak chaotic layer is{,
however,} still present, {but} the 1:1 resonance is confined to
the center of the grid producing multiple stability islands. The
1:2 resonance take{s} over almost the entire phase plane in
Figure~\ref{gridsE}(e) where $E_{\rm J} = 593$. In this case we
should note that apart {from} the 1:2 type resonance{,} the 1:2
type b (corresponding to figure-eight orbits) emerges at the outer
parts of the phase plane. Things are quite different in
Figure~\ref{gridsE}(f) where $E_{\rm J} = 780$ and $x_{\max} = 4$
kpc. We see a small decrease {in} the extent of the 1:2 (type a)
islands due to the increase {in} the amount of box orbits at the
outer parts of the phase plane, while 1:2 type b orbits disappear.
At the center of the grid we{ can} distinguish a well-formed 1:1
stability island, while in the neighborhood there is a mixture of
delocalized initial conditions corresponding to chaotic, 1:1 and
higher resonant orbits.

Looking at Figure~\ref{gridsE}(a)--(f) 
 we see
that{,} as the value of the energy increases, in other words we
study the motion of stars moving at larger distances from the
galactic center, the 1:2 resonant orbits dominate the vast
majority of the phase plane. Here we should notice that according
to SM the value of the strength of the bar is $\alpha = 4${,} so
the potential of the harmonic oscillator used for the description
of the bar is at resonance; the 1:2 resonance to be more precise.
Thus, a natural and fair question arises: what happens to the
phase plane if we change the value of the strength of the bar? To
give an answer to this question, we chose the value $\alpha = 9$
(which{ also} corresponds to integer resonance; 1:3) and
reconstructed the grid of initial conditions shown in
Figure~\ref{gridsE}(e). Our results are given in Figure~\ref{R13}.
 Now, the harmonic oscillator is at the 1:3 resonance and
it can be seen that the 1:3 resonance prevails, while the 1:2
resonant orbits define two small stability islands confined to the
center of the grid. This is because the model assumes {a} constant
ratio of $x$ and $y$ frequencies throughout, which however
generally is not the case. Therefore, we conclude that for high
values of the orbital energy corresponding to star orbits moving
sufficiently far from the central nucleus ($x_{\max} > 1.5$~kpc),
the influence of the bar prevails over that of the central
nucleus.

\begin{figure*}[!tH]
\centering


\vs \resizebox{0.88\hsize}{!}{\includegraphics{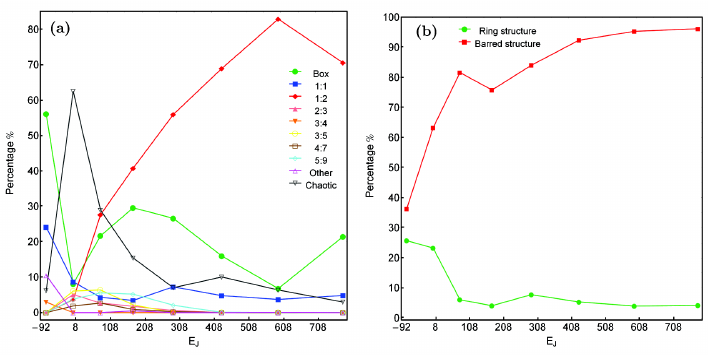}}

\caption{ \baselineskip 3.6mm Evolution of the percentages of
(a-{\it left}): the different types of orbits in our barred galaxy
model and (b-{\it right}): the types of regular orbits supporting
the formation of nuclear rings {or a} barred structure, when
varying the value of the Jacobi integral $E_{\rm J}$.}
\label{percsE}

\vs\vs

\vs \centering

\vs \resizebox{0.85\hsize}{!}{\includegraphics{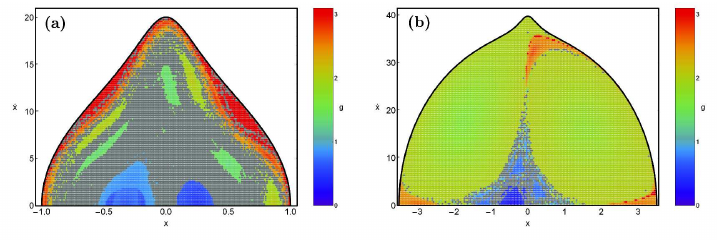}}

\caption{\baselineskip 3.6mm Grids of initial conditions
$(x_0,\dot{x_0})$ when (a-{\it left}): $E_{\rm J} = 1$ and (b-{\it
right}): $E_{\rm J} = 593$. Each point is colored according to the
value of the ratio $g$, thus distinguishing between regular orbits
supporting ring formation $(0.5 < g < 1.5)$ and{ a} barred
structure $(g \geq 2)$.} \label{pRB}
\end{figure*}

In the following Figure~\ref{percsE}(a) we present the evolution
of the resulting percentages of chaotic and regular orbits for the
barred galaxy model as the value of the energy $E_{\rm J}$ varies.
We observe that when the motion of stars is at low energies, i.e.,
orbits which move very close to the galactic center, {they are}
highly regular, {with} box and 1:1 resonant orbits {being }the
most populated families. The largest amount of chaotic orbits
(around 65\%) is observed when the value of energy is near zero,
{but} with increasing $E_{\rm J}$ their rate drops rapidly and at
higher energy values it eventually vanishes. The percentage of box
orbits displays strong fluctuations with sudden drops and peaks.
The rate of the 1:1 resonant orbits on the other hand, for $E_{\rm
J} > 55${,} evolves almost monotonically{,} keeping a constant
value around 5\%. It is also seen that the percentage of 1:2 (type
a) resonant orbits starts to grow sharply as soon as the energy
grows and when $E_{\rm J} > 100$ the 1:2 (type a) family is the
dominant type.
 In fact, we see that for high values of the
Jacobi integral ($E_{\rm J} = 600$) the 1:2 resonant orbits take
over more than 80\% of the entire phase plane, although for larger
values of energy their rates exhibit a small decrease due to the
simultaneous increase of the box orbits. All the other
resonant families seem to be immune to the increase {in} energy
since their percentages are almost unperturbed throughout; all these
families practically disappear when $E_{\rm J}
> 400$. Takin{g} into consideration all the above-mentioned analysis,
we may conclude that in barred galaxy models the value of energy
{mostly }affects {the} chaotic, box, 1:1 and 1:2 (type a) resonant orbits.
We would like to point out that the 1:2 type b resonant orbits are
{only }present in barred galaxy models with high enough energy.

Of particular interest is to interpret the evolution of the
percentages of regular orbits supporting{ a} ring {or} barred
structure as a function of the Jacobi integral $E_{\rm J}$. Our
numerical results are given in Figure~\ref{percsE}(b) where it can
be seen that {the} barred structure is {always }more favored. To
be more precise, for low energies ($E_{\rm J} < 20$) only about
25\% of the total regular orbits support the formation of nuclear
rings, while for larger values of energy the rate drops{ to}
around 5\% and remains there throughout. On the contrary, we see
that the percentage of regular orbits supporting the bar grows
rapidly with increasing energy, although {in} high energy models
($E_{\rm J} > 410$) their percentage seems to saturate around
90\%. Summarizing, low energy models support the formation of both
nuclear rings and bars, while high energy models are {only
}favored {for} barred structure{s}.

Figure~\ref{pRB}(a)--(b) 
 shows 
 another {perspective
related to} the grids of Figure~\ref{gridsE}(b) and (e). Here,
each initial condition is colored according to the value of the
ratio $g$, thus distinguishing between regular orbits supporting
ring formation $(0.5 < g < 1.5)$ and {a }barred structure $(g \geq
2)$. It becomes evident that the 1:1 resonant orbits indeed{
support} the formation of nuclear rings, while the 1:2 (type a and
b) resonant orbits support the barred structure of the galaxy.

\section{Conclusions}
\label{disc}

In this work, we used an analytic galactic gravitational model which
embraces the general features of a barred galaxy containing a
spherical, dense and massive nucleus. The choice of the model
potential for the description of the bar was made mainly taking into
account{ the fact} that near the center of a galaxy the motion of stars can be
approximated by harmonic oscillations. Our aim was to investigate
how the basic parameters of the Hamiltonian system influence the
level of chaos and also the distribution of regular families in our
barred galaxy model. Our results strongly suggest that both
the level of chaos and the distribution of regular families are
indeed very dependent on the parameters of the galaxy. We
believe that the presented outcomes can provide interesting
information regarding the structure and properties of barred
galaxies.

In our research, we chose to investigate the influence of four
basic quantities {that are part }of the galactic model, namely the
strength of the bar, the scale length of the nucleus, the angular
velocity of the bar and the value of the total orbital energy
(Jacobi constant). We decided not to explore the influence of the
mass of the nucleus for two main reasons: (i) it has been
extensively studied in earlier works \citep[see,
e.g.][]{HN90,HPN93,Z12a,ZC13} and (ii) in this model the mass of
the nucleus significantly{ affects} the size of the grid (in other
words the values of $x_{\max}$ and $\dot{x_{\max}}$){,} so it was
impossible to set a constant energy level and then vary the mass
of the nucleus. We also tried to find out which regular orbits
support the barred structure of the galaxy and which support the
formation of nuclear rings, using the $g$ value as the only
criterion for this task. The main results of our research can be
summarized as follows:
\begin{enumerate}

 \item[(1)] Perhaps the most important finding of our research is the fact that the traditional
 $x_1$ orbital family does not {always }dominate the structures of all {2D} barred galaxy models{,} thus verifying similar outcomes \citep[see, e.g.][]{BT01,SPA02b,KP05}. Indeed, we have presented numerical evidence that several resonant orbits which are not related {to} the
 $x_1$ family can support the bar.
 \item[(2)] In our barred galaxy model{,} several types of regular orbits exist, {but} there are also extended chaotic domains separating the areas of regularity. In particular, a large variety of resonant orbits (i.e. 1:1, 1:2, 1:3, 2:3, 3:4, 3:5, 4:7, 5:7, 5:8, 5:9 and higher resonant orbits) are present, thus making the orbital structure {richer}. Here we must clarify that by the term ``higher resonant orbits" we refer to resonant orbits with a rational quotient of frequencies made from integers $> 5$, which of course do not belong to the main families.
 \item[(3)] It was found that in barred galaxy models the strength of the bar $\alpha$ {mostly }influences box, 1:1, 1:2 and chaotic orbits, turning a large portion of 1:1 and box orbits into 1:2 and chaotic orbits as the bar becomes stronger, or in other words, as the value of $\alpha$ increases. As expected, galaxy models with relatively strong bars do not favor the formation of nuclear rings.
 \item[(4)] Increasing the scale length of the nucleus (in other words the nucleus becomes less concentrated and dense) {mainly }turns chaotic orbits into box orbits, while resonant orbits are less affected. Dense nuclei {only }favor the formation of nuclear rings, {but} when the central nucleus is sparse enough about 40\% of the total types of regular orbits support the barred structure and only 20\% of them the nuclear rings.
 \item[(5)] As the bar gains speed, different types of regular orbits become chaotic{,} occupying more than 70\% of the entire phase plane. We found that slow{ly} rotat{ing} bars support the formation of nuclear rings, while fast rotat{ing} bars{ mainly} favor the barred structure.
 \item[(6)] A strong correlation between the value of the energy and the percentages of chaotic, box, 1:1 and 1:2 resonant orbits was found to exist. Moreover, low energy models support the formation of both nuclear rings and bars, while high energy models{ only} favor {the} barred structure.
 \end{enumerate}

We consider the present results as an initial effort and also a
promising step in the task of understanding the orbital structure of
barred galaxies. Taking into account that our outcomes are
encouraging, it is in our future plans to utilize a logarithmic
potential for describing the properties of the bar, thus expanding
our investigation in global motion as well as into three dimensions,
exploring how the basic parameters influence the nature of the
3D orbits. Furthermore, of particular interest
would be to reveal the complete network of periodic orbits, thus
shedding some light {on} the evolution of periodic
orbits as well as their stability when varying all the available
parameters of the galactic model.

\begin{acknowledgements}

The authors would like to express their warmest thanks to the
anonymous referee for the careful reading of the manuscript and
for all the apt suggestions and comments which allowed us to
significantly{ improve} both the quality and the clarity of our
paper.
\end{acknowledgements}

\vs
\appendix
\section{Sticky orbits and numerical integration time}

\begin{figure}

\vs \centering
\includegraphics[width=0.9\hsize]{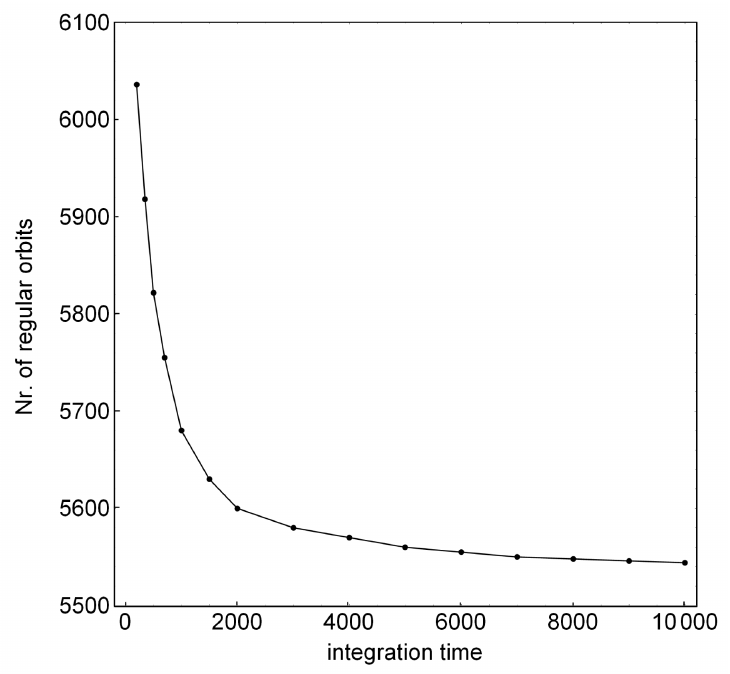}

\caption{\baselineskip 3.6mm  Evolution of the total number of
orbits classified as regular in the SM model, using the SALI chaos
indicator, as a function of the total time of the numerical
integration. The threshold value of SALI was fixed at $10^{-7}$.
\label{tfin}}
\end{figure}

\begin{table}
   \centering

\begin{minipage}{80mm}

\caption{\baselineskip 3.6mm Total number of orbits identified as
chaotic at $10^2$ $(N)$ and at $10^{4}$ $(N')$ time units of
numerical integration for all studied models.
   \label{table0}}\end{minipage}

   \fns\tabcolsep 2.5mm
   \begin{tabular}{lcccccccc}
      \hline
      $\alpha$ & $c_{\rm n}$ & $\Omega_{\rm b}$ & $E_{\rm J}$ & $N$ & $N'$ & Error (\%) \\
      \hline
      1 & 0.25 & 1 & 1.5 & 0 & 0 & 0 \\
      2 & 0.25 & 1 & 1.5 & 1575 & 1651 & 4.6 \\
      3 & 0.25 & 1 & 1.5 & 6459 & 6749 & 4.3 \\
      4 & 0.25 & 1 & 1.5 & 8971 & 9463 & 5.2 \\
      5 & 0.25 & 1 & 1.5 & 10498 & 11253 & 6.7 \\
      6 & 0.25 & 1 & 1.5 & 10506 & 11106 & 5.4 \\
      7 & 0.25 & 1 & 1.5 & 9972 & 10207 & 2.3 \\
      8 & 0.25 & 1 & 1.5 & 5363 & 5592 & 4.1 \\
      9 & 0.25 & 1 & 1.5 & 8409 & 8851 & 5.0 \\
     10 & 0.25 & 1 & 1.5 & 7628 & 8108 & 5.9 \\
      4 & 0.05 & 1 & 1.5 & 12905 & 13802 & 6.5 \\
      4 & 0.10 & 1 & 1.5 & 12997 & 13511 & 3.8 \\
      4 & 0.15 & 1 & 1.5 & 12822 & 13509 & 5.1 \\
      4 & 0.20 & 1 & 1.5 & 12027 & 12607 & 4.6 \\
      4 & 0.25 & 1 & 1.5 & 8971 & 9463 & 5.2 \\
      4 & 0.30 & 1 & 1.5 & 5083 & 5256 & 3.3 \\
      4 & 0.35 & 1 & 1.5 & 146 & 148 & 1.5 \\
      4 & 0.40 & 1 & 1.5 & 0 & 0 & 0 \\
      4 & 0.45 & 1 & 1.5 & 0 & 0 & 0 \\
      4 & 0.50 & 1 & 1.5 & 0 & 0 & 0 \\
      4 & 0.25 & 0.00 & 1.5 & 4123 & 4203 & 1.9 \\
      4 & 0.25 & 0.25 & 1.5 & 5072 & 5108 & 0.7 \\
      4 & 0.25 & 0.50 & 1.5 & 6949 & 7201 & 3.5 \\
      4 & 0.25 & 0.75 & 1.5 & 8358 & 8556 & 2.3 \\
      4 & 0.25 & 1.00 & 1.5 & 8971 & 9463 & 5.2 \\
      4 & 0.25 & 1.25 & 1.5 & 9785 & 10056 & 2.7 \\
      4 & 0.25 & 1.50 & 1.5 & 10062 & 10648 & 5.5 \\
      4 & 0.25 & 1.75 & 1.5 & 10295 & 10952 & 6.0 \\
      4 & 0.25 & 2.00 & 1.5 & 10502 & 10950 & 4.1 \\
      4 & 0.25 & 2.25 & 1.5 & 10412 & 10812 & 3.7 \\
      4 & 0.25 & 2.50 & 1.5 & 10039 & 10657 & 5.8 \\
      4 & 0.25 & 1 & --77 & 918 & 931 & 1.4 \\
      4 & 0.25 & 1 &  1 & 8866 & 9452 & 6.2 \\
      4 & 0.25 & 1 & 80 & 4152 & 4351 & 4.6 \\
      4 & 0.25 & 1 & 174 & 2339 & 2392 & 2.2 \\
      4 & 0.25 & 1 & 290 & 1014 & 1055 & 3.9 \\
      4 & 0.25 & 1 & 430 & 1495 & 1512 & 1.1 \\
      4 & 0.25 & 1 & 593 & 976 & 972 & 2.8 \\
      4 & 0.25 & 1 & 780 & 451 & 455 & 0.9 \\
      \hline
   \end{tabular}
\end{table}

Usually, when investigating the regular or chaotic nature of
orbits in galactic potentials, we try to keep the integration time
as long as 1 Hubble time because this allows {us }to relate the
calculations to our Universe. However, in previous research
\citep{ZC13,ZCar14} we {demonstrated} that{ when} using short
integration time intervals (equal to about 1 to 10 Hubble
times){,} a non-negligible number of chaotic orbits were
misclassified as regular by several chaos indicators. This
phenomenon is also true in the present case.

Figure~\ref{tfin} 
 shows{,} for the set of orbits
{in} the SM model, how the number of regular orbits shifted along
with the time span of the orbital integration. As a reference,
Table~\ref{table0} 
  shows{,} for each of the models
that was studied in Section~\ref{numres}, the total number of
orbits identified as chaotic at $10^2$ and also at $10^{4}$ time
units of numerical integration. It is evident from
Table~\ref{table0}, that{ when} using integration time much longer
than 1 Hubble time{,} the number of misclassified orbits in every
model has been considerably reduced (the maximum relative error
that was measured was about 7\%). Nevertheless, even this extended
numerical integration {does} not {completely }solve the problem;
in fact, there will always be (in a non integrable potential)
sticky orbits which behave as regular ones during arbitrarily
large times, rendering any attempt to develop an algorithm which
finds them all in a short time hopeless.


\begin{thebibliography}{99}
\small \setlength{\itemindent}{-3mm} \setlength{\itemsep}{-0.5mm}
\setlength{\baselineskip}{4.6mm}



\bibitem[{Athanassoula}(1984)]{A84}
{Athanassoula}, E. 1984, \physrep, 114, 319

\bibitem[{Athanassoula}(1992)]{A92}
{Athanassoula}, E. 1992, \mnras, 259, 345

\bibitem[{Athanassoula} {et~al.}(1983)]{ABMP83}
{Athanassoula}, E., {Bienayme}, O., {Martinet}, L., \& {Pfenniger}, D. 1983,
  \aap, 127, 349

\bibitem[{Barnes} \& {Tohline}(2001)]{BT01}
{Barnes}, E.~I., \& {Tohline}, J.~E. 2001, \apj, 551, 80

\bibitem[{Benedict} {et~al.}(2002)]{BHJKS02}
{Benedict}, G.~F., {Howell}, D.~A., {J{\o}rgensen}, I., {Kenney}, J.~D.~P., \&
  {Smith}, B.~J. 2002, \aj, 123, 1411

\bibitem[{Binney} \& {Spergel}(1982)]{BS82}
{Binney}, J., \& {Spergel}, D. 1982, \apj, 252, 308

\bibitem[{Binney} \& {Spergel}(1984)]{BS84}
{Binney}, J., \& {Spergel}, D. 1984, \mnras, 206, 159

\bibitem[{Binney} \& {Tremaine}(2008)]{BT08}
{Binney}, J., \& {Tremaine}, S. 2008, {Galactic Dynamics: Second Edition}
  (Princeton: Princeton Univ. Press)

\bibitem[{Bountis} {et~al.}(2012)]{BMA12}
{Bountis}, T., {Manos}, T., \& {Antonopoulos}, C. 2012, Celestial Mechanics and
  Dynamical Astronomy, 113, 63

\bibitem[{Buta} {et~al.}(2000)]{BTBC00}
{Buta}, R., {Treuthardt}, P.~M., {Byrd}, G.~G., \& {Crocker}, D.~A. 2000, \aj,
  120, 1289

\bibitem[{Caranicolas} \& {Barbanis}(1982)]{CB82}
{Caranicolas}, N., \& {Barbanis}, B. 1982, \aap, 114, 360

\bibitem[{Caranicolas}(1998)]{C98}
{Caranicolas}, N.~D. 1998, \aap, 332, 88

\bibitem[{Caranicolas} \& {Karanis}(1998)]{CK98}
{Caranicolas}, N.~D., \& {Karanis}, G.~I. 1998, \apss, 259, 45

\bibitem[{Caranicolas} \& {Papadopoulos}(2005)]{CP05}
{Caranicolas}, N.~D., \& {Papadopoulos}, N.~J. 2005, Baltic Astronomy, 14, 535

\bibitem[{Caranicolas} \& {Papadopoulos}(2007)]{CP07}
{Caranicolas}, N.~D., \& {Papadopoulos}, N.~J. 2007, Astronomische Nachrichten,
  328, 556

\bibitem[{Caranicolas} \& {Zotos}(2010)]{CZ10}
{Caranicolas}, N.~D., \& {Zotos}, E.~E. 2010, \na, 15, 427

\bibitem[{Caranicolas} \& {Zotos}(2013)]{CZ13}
{Caranicolas}, N.~D., \& {Zotos}, E.~E. 2013, \pasa, 30, 49

\bibitem[{Carpintero} \& {Aguilar}(1998)]{CA98}
{Carpintero}, D.~D., \& {Aguilar}, L.~A. 1998, \mnras, 298, 1

\bibitem[{Carpintero} \& {Muzzio}(2012)]{CM12}
{Carpintero}, D.~D., \& {Muzzio}, J.~C. 2012, Celestial Mechanics and Dynamical
  Astronomy, 112, 107

\bibitem[{Combes} {et~al.}(1990)]{CDFP90}
{Combes}, F., {Debbasch}, F., {Friedli}, D., \& {Pfenniger}, D. 1990, \aap,
  233, 82

\bibitem[{Comer{\'o}n} {et~al.}(2010)]{CKB10}
{Comer{\'o}n}, S., {Knapen}, J.~H., {Beckman}, J.~E., {et~al.} 2010, \mnras,
  402, 2462

\bibitem[{Contopoulos} \& {Barbanis}(1985)]{CB85}
{Contopoulos}, G., \& {Barbanis}, B. 1985, \aap, 153, 44

\bibitem[{Contopoulos} \& {Grosbol}(1989)]{CG89}
{Contopoulos}, G., \& {Grosbol}, P. 1989, \aapr, 1, 261

\bibitem[{Contopoulos} \& {Magnenat}(1985)]{CM85}
{Contopoulos}, G., \& {Magnenat}, P. 1985, Celestial Mechanics,
37,~387

\bibitem[{Contopoulos} \& {Mertzanides}(1977)]{CM77}
{Contopoulos}, G., \& {Mertzanides}, C. 1977, \aap, 61, 477

\bibitem[{Contopoulos} \& {Papayannopoulos}(1980)]{CP80}
{Contopoulos}, G., \& {Papayannopoulos}, T. 1980, \aap, 92, 33

\bibitem[{Englmaier} \& {Gerhard}(1997)]{EG97}
{Englmaier}, P., \& {Gerhard}, O. 1997, \mnras, 287, 57

\bibitem[{Eskridge} {et~al.}(2000)]{Ee00}
{Eskridge}, P.~B., {Frogel}, J.~A., {Pogge}, R.~W., {et~al.} 2000,
\aj, 119,~536

\bibitem[Ferrers(1877)]{F77}
Ferrers, N.~M. 1877, The Quarterly Journal of Pure and Applied Mathematics, 14,
  1

\bibitem[{Friedli} \& {Martinet}(1993)]{FM93}
{Friedli}, D., \& {Martinet}, L. 1993, \aap, 277, 27

\bibitem[{Hasan} \& {Norman}(1990)]{HN90}
{Hasan}, H., \& {Norman}, C. 1990, \apj, 361, 69

\bibitem[{Hasan} {et~al.}(1993)]{HPN93}
{Hasan}, H., {Pfenniger}, D., \& {Norman}, C. 1993, \apj, 409, 91

\bibitem[{Hsieh} {et~al.}(2011)]{HMLHOW11}
{Hsieh}, P.-Y., {Matsushita}, S., {Liu}, G., {et~al.} 2011, \apj, 736, 129

\bibitem[{Kaufmann} \& {Contopoulos}(1996)]{KC96}
{Kaufmann}, D.~E., \& {Contopoulos}, G. 1996, \aap, 309, 381

\bibitem[{Kaufmann} \& {Patsis}(2005)]{KP05}
{Kaufmann}, D.~E., \& {Patsis}, P.~A. 2005, \apj, 624, 693

\bibitem[{Kim} {et~al.}(2012{\natexlab{a}})]{KSK12}
{Kim}, W.-T., {Seo}, W.-Y., \& {Kim}, Y. 2012{\natexlab{a}}, \apj, 758, 14

\bibitem[{Kim} {et~al.}(2012{\natexlab{b}})]{KSSYT12}
{Kim}, W.-T., {Seo}, W.-Y., {Stone}, J.~M., {Yoon}, D., \& {Teuben}, P.~J.
  2012{\natexlab{b}}, \apj, 747, 60

\bibitem[{Knapen} {et~al.}(1995)]{KBHSd95}
{Knapen}, J.~H., {Beckman}, J.~E., {Heller}, C.~H., {Shlosman}, I., \& {de
  Jong}, R.~S. 1995, \apj, 454, 623

\bibitem[{Knapen}(2004)]{K04}
{Knapen}, J.~H. 2004, in Astrophysics and Space Science Library, 
 319,
  Penetrating Bars Through Masks of Cosmic Dust, ed. D.~L. {Block},
  I.~{Puerari}, K.~C. {Freeman}, R.~{Groess}, \& E.~K. {Block}, 189

\bibitem[{Kormendy} \& {Kennicutt}(2004)]{KK04}
{Kormendy}, J., \& {Kennicutt}, Jr., R.~C. 2004, \araa, 42, 603

\bibitem[{Lees} \& {Schwarzschild}(1992)]{LS92}
{Lees}, J.~F., \& {Schwarzschild}, M. 1992, \apj, 384, 491

\bibitem[{Maciejewski} {et~al.}(2002)]{MTSS02}
{Maciejewski}, W., {Teuben}, P.~J., {Sparke}, L.~S., \& {Stone}, J.~M. 2002,
  \mnras, 329, 502

\bibitem[{Manos} \& {Athanassoula}(2011)]{MA11}
{Manos}, T., \& {Athanassoula}, E. 2011, \mnras, 415, 629

\bibitem[{Manos} {et~al.}(2013)]{MBS13}
{Manos}, T., {Bountis}, T., \& {Skokos}, C. 2013, Journal of Physics A
  Mathematical General, 46, 254017

\bibitem[{Marinova} \& {Jogee}(2007)]{MJ07}
{Marinova}, I., \& {Jogee}, S. 2007, \apj, 659, 1176

\bibitem[{Mazzuca} {et~al.}(2008)]{MKVR08}
{Mazzuca}, L.~M., {Knapen}, J.~H., {Veilleux}, S., \& {Regan}, M.~W. 2008,
  \apjs, 174, 337

\bibitem[{Mu{\~n}oz-Tu{\~n}{\'o}n} {et~al.}(2004)]{MCA04}
{Mu{\~n}oz-Tu{\~n}{\'o}n}, C., {Caon}, N., \& {Aguerri}, J.~A.~L. 2004, \aj,
  127,~58

\bibitem[{Muzzio} {et~al.}(2005)]{MCW05}
{Muzzio}, J.~C., {Carpintero}, D.~D., \& {Wachlin}, F.~C. 2005, Celestial
  Mechanics and Dynamical Astronomy, 91, 173

\bibitem[{Olle} \& {Pfenniger}(1998)]{OP98}
{Olle}, M., \& {Pfenniger}, D. 1998, \aap, 334, 829

\bibitem[{Pfenniger}(1984)]{P84}
{Pfenniger}, D. 1984, \aap, 134, 373

\bibitem[{Pfenniger}(1996)]{P96}
{Pfenniger}, D. 1996, in Astronomical Society of the Pacific Conference Series,
  91, IAU Colloq. 157: Barred Galaxies, ed. R.~{Buta}, D.~A. {Crocker}, \&
  B.~G. {Elmegreen}, 273

\bibitem[{Pichardo} {et~al.}(2004)]{PMM04}
{Pichardo}, B., {Martos}, M., \& {Moreno}, E. 2004, \apj, 609, 144

\bibitem[{Piner} {et~al.}(1995)]{PST95}
{Piner}, B.~G., {Stone}, J.~M., \& {Teuben}, P.~J. 1995, \apj, 449, 508

\bibitem[{Press} {et~al.}(1992)]{PTVF92}
{Press}, W.~H., {Teukolsky}, S.~A., {Vetterling}, W.~T., \&
{Flannery}, B.~P.
  1992, {Numerical Pecipes in FORTRAN.  The Art of \newpage  Scientific Computing}
  (Cambridge: Cambridge Univ. Press)

\bibitem[{Regan} \& {Teuben}(2003)]{RT03}
{Regan}, M.~W., \& {Teuben}, P. 2003, \apj, 582, 723

\bibitem[{Sandstrom} {et~al.}(2010)]{Se10}
{Sandstrom}, K., {Krause}, O., {Linz}, H., {et~al.} 2010, \aap, 518, L59

\bibitem[{Sellwood} \& {Wilkinson}(1993)]{SW93}
{Sellwood}, J.~A., \& {Wilkinson}, A. 1993, Reports on Progress in Physics, 56,
  173

\bibitem[{Sheth} {et~al.}(2003)]{SRSS03}
{Sheth}, K., {Regan}, M.~W., {Scoville}, N.~Z., \& {Strubbe}, L.~E. 2003,
  \apjl, 592, L13
\bibitem[{Sheth} {et~al.}(2005)]{SVRTT05}
{Sheth}, K., {Vogel}, S.~N., {Regan}, M.~W., {Thornley}, M.~D., \& {Teuben},
  P.~J. 2005, \apj, 632, 217




\bibitem[{Skokos}(2001)]{S01}
{Skokos}, C. 2001, Journal of Physics A Mathematical General, 34, 10029

\bibitem[{Skokos} {et~al.}(2002{\natexlab{a}})]{SPA02a}
{Skokos}, C., {Patsis}, P.~A., \& {Athanassoula}, E. 2002{\natexlab{a}},
  \mnras, 333, 847

\bibitem[{Skokos} {et~al.}(2002{\natexlab{b}})]{SPA02b}
{Skokos}, C., {Patsis}, P.~A., \& {Athanassoula}, E. 2002{\natexlab{b}},
  \mnras, 333, 861



\bibitem[{Thakur} {et~al.}(2009)]{TAJ09}
{Thakur}, P., {Ann}, H.~B., \& {Jiang}, I.-G. 2009, \apj, 693, 586

\bibitem[{{\v S}idlichovsk{\'y}} \& {Nesvorn{\'y}}(1996)]{SN96}
{{\v S}idlichovsk{\'y}}, M., \& {Nesvorn{\'y}}, D. 1996, Celestial Mechanics
  and Dynamical Astronomy, 65, 137

\bibitem[{Zotos}(2011)]{Z11}
{Zotos}, E.~E. 2011, Baltic Astronomy, 20, 339

\bibitem[{Zotos}(2012a)]{Z12a}
{Zotos}, E.~E. 2012a, \na, 17, 576

\bibitem[{Zotos}(2012b)]{Z12b}
Zotos, E.~E. 2012b, Nonlinear Dynamics, 69, 2041

\bibitem[Zotos(2013)]{Z13}
Zotos, E.~E. 2013, Nonlinear Dynamics, 73, 931

\bibitem[{Zotos} \& {Caranicolas}(2013)]{ZCar13}
{Zotos}, E.~E., \& {Caranicolas}, N.~D. 2013, \aap, 560, A110

\bibitem[Zotos \& Caranicolas(2014)]{ZCar14}
Zotos, E.~E., \& Caranicolas, N.~D. 2014, Nonlinear Dynamics, 76, 323

\bibitem[{Zotos} \& {Carpintero}(2013)]{ZC13}
{Zotos}, E.~E., \& {Carpintero}, D.~D. 2013, Celestial Mechanics and Dynamical
  Astronomy, 116, 417

\end{thebibliography}
\end{document}